\documentclass[journal,10pt]{IEEEtran} 

\usepackage{amsmath}
\usepackage{amsthm}

\usepackage{breqn}
\usepackage{newtxtext} 
\usepackage{mathptmx} 
\usepackage{mathrsfs}
\usepackage{cite}
\usepackage{graphicx}
\usepackage[usenames, dvipsnames]{color}
\usepackage[english]{babel}

\newcommand{\Tabref}[1]{Table~\ref{#1}}
\newcommand{\Figref}[1]{Fig.~\ref{#1}}

\newcommand{\Sectref}[1]{Section~\ref{#1}}

\newcommand\ceil[1]{\lceil#1\rceil}

\newtheorem{theorem}{Theorem}[section]

\begin{document}
\title{ \LARGE \bf Fast Node Cardinality Estimation and Cognitive MAC Protocol Design for Heterogeneous Machine-to-Machine Networks}
\author {Sachin Kadam, Chaitanya S. Raut, Amandeep Meena, and Gaurav S. Kasbekar}
\maketitle{}
{\renewcommand{\thefootnote}{} \footnotetext{S. Kadam and G. Kasbekar are with Department of Electrical Engineering, Indian Institute of Technology (IIT) Bombay, Mumbai, India, C. Raut is with Druva Software, Pune, India and A. Meena is with Toshiba Software Pvt. Ltd., Bengaluru, India. Their email addresses are \{sachink, gskasbekar\}@ee.iitb.ac.in, chaitanya.raut@druva.com, and  amandeep.meena@toshiba-tsip.com respectively. C. Raut and A. Meena worked on this research while they were with IIT Bombay.
\par A preliminary version of this paper~\cite{GlobecomPaper} was presented at the IEEE GLOBECOM 2017 conference and was published in its proceedings (DOI: 10.1109/GLOCOM.2017.8254618).
}}

\begin{abstract}
Machine-to-Machine (M2M) networks are an emerging technology with
applications in numerous areas including smart grids, smart cities,
vehicular telematics, and healthcare.  In this paper, we design two
estimation protocols for rapidly obtaining separate estimates of the
number of active nodes of each traffic type in a heterogeneous M2M network
with $T$ types of M2M nodes (e.g., those that send emergency, periodic,
normal type data etc), where $T \geq 2$ is an arbitrary integer. One of
these protocols, Method I, is a simple scheme, and the other, Method II,
is more sophisticated and performs better than Method I. Also, we design a
medium access control (MAC) protocol that supports multi-channel operation
for a heterogeneous M2M network with an arbitrary number of types of M2M
nodes, operating as a secondary network using Cognitive Radio technology.
Our Cognitive MAC protocol uses the proposed node cardinality estimation
protocols to rapidly estimate the number of active nodes of each type in
every time frame; these estimates are used to find the optimal contention
probabilities to be used in the MAC protocol. We compute a closed form
expression for the expected number of time slots required by Method I to
execute as well as a simple upper bound on it. Also, we mathematically
analyze the performance of the Cognitive MAC protocol and obtain
expressions for the expected number of successful contentions per frame
and the expected amount of energy consumed. Finally, we evaluate the
performances of our proposed estimation protocols and Cognitive MAC
protocol using simulations.
\end{abstract}

\section{Introduction}\label{Intro}

Machine-to-Machine (M2M) communications is an emerging technology, in which data generation, processing and transmission is done with minimal human intervention~\cite{wu2011m2m}. M2M networks have applications in numerous areas including smart grids, smart cities, vehicular telematics, healthcare, industrial automation, security and public safety~\cite{wu2011m2m,zhang2012cognitive,rajandekar2015survey}. It is challenging to design medium access control (MAC) protocols for M2M networks due to their unique characteristics such as limited access to energy sources (most M2M devices are battery operated), need to provide network access to a very large number of devices, the fact that the Quality of Service (QoS) requirements of M2M devices differ from those of Human-to-Human (H2H) communications and are also different for different M2M devices~\footnote{For example, some M2M nodes need to transmit data (e.g., smart meter readings) periodically, some need to send emergency or alarm messages (e.g., in healthcare and security applications), some need to transmit normal data traffic and some need to reliably transmit  data packets (e.g., in remote payment gateway systems)~\cite{wu2011m2m,rajandekar2015survey,3gpp}.} etc~\cite{wu2011m2m,rajandekar2015survey,3gpp}. 

Several wireless technologies such as Bluetooth, Wi-Fi, ZigBee and cellular networks including LTE-Advanced and 802.16 are potential candidates for enabling M2M communications; however, these technologies have some shortcomings~\footnote{Specifically, Wi-Fi has high power consumption, due to which it is not suitable for battery operated M2M devices, and Bluetooth has high latency when the number of devices is large, as is the case in M2M networks~\cite{aijaz2015prma}. ZigBee operates on unlicensed bands and is prone to interference from Wi-Fi networks and other equipment (e.g., microwave ovens) that use those bands~\cite{aijaz2015prma,zhang2012cognitive}. Due to the high demand for H2H communication services such as voice, video, emails etc, only a limited amount of radio spectrum is available with cellular operators to support M2M communications~\cite{zhang2012cognitive}.}~\cite{aijaz2015prma}. 
Cognitive Radio technology is a promising alternative to the above wireless technologies for enabling M2M communications~\cite{zhang2012cognitive}. Cognitive Radio Networks (CRNs) have emerged as a promising solution to alleviate the artificial spectrum scarcity (wherein most of the usable radio spectrum is allocated, but underutilized) caused by the traditional spectrum regulation policy of assigning \emph{exclusive} licenses to users to operate their networks in different geographical regions~\cite{akyildiz2006next}. In CRNs, there are two types of spectrum users-- primary users (PUs), which have prioritized access to channels, and secondary users (SUs) that detect and use spectrum holes, i.e., chunks of spectrum that are currently not in use by the PUs~\cite{akyildiz2006next}. Operating an M2M network as a secondary network using Cognitive Radio technology has the advantage that a large amount of spectrum, which is allocated to other users, but underutilized, becomes available for M2M communications~\cite{aijaz2015prma}. However, this requires the design of efficient \emph{Cognitive MAC protocols} in order to provide channel access to an extremely large number of M2M devices, while satisfying the unique service requirements of M2M applications described in the first paragraph of this section, as well as ensuring avoidance of interference to PUs. 
The design of a Cognitive MAC protocol~\emph{that supports multi-channel operation} involves addressing additional challenges~\cite{de2012survey} including achieving coordination among nodes~\footnote{Note that for two nodes to be able to exchange data, both must have their wireless transceiver tuned to a common channel at a time.}, overcoming the multi-channel hidden terminal problem~\cite{so2004multi}, and balancing the traffic load of the secondary (M2M) nodes over the free channels in real-time~\cite{cordeiro2007c}.
In this paper, \emph{we design a Cognitive MAC protocol for M2M networks that  overcomes the above challenges}.

Now, consider an M2M network in which a large number of M2M devices intermittently transmit some information (e.g., smart meter readings, information collected by sensors) to a base station (BS). In any given time frame, the BS is unaware of the number of \textit{active} nodes, i.e., those that need to transmit some data to the BS in the current frame. There is a need to rapidly estimate the number of active nodes since this estimate can be used to determine the optimal values of various parameters of the MAC protocol such as contention probabilities and the amounts of time to be used for contention and for data transmission in the current frame. For example, recall that for the Slotted ALOHA protocol, the optimal contention probability is the reciprocal of the number of active nodes~\cite{bertsekas1992data}. Also, in~\cite{wu2013fasa, park2014enhancement, hsu2013adaptive}, active node cardinality estimation is performed and using the estimates obtained, the contention probabilities that maximize the throughput of their respective MAC protocols for M2M networks are determined.  

In a \emph{heterogeneous} M2M network, i.e., one in which different types of nodes are present (e.g., those that send emergency, periodic and normal type data), we need to obtain \textit{separate} estimates of the number of active nodes of each traffic type. In prior work, several protocols have been designed~\cite{wu2013fasa, park2014enhancement, hsu2013adaptive, Est_CSMA, Congestion_Liu} to estimate the number of active nodes in a \textit{homogeneous} M2M network (see Section~\ref{ReWo}). However, to the best of our knowledge, so far \textit{no estimation protocol has been designed for obtaining separate estimates of the number of active nodes of each traffic type in a heterogeneous M2M network}. Note that executing a node cardinality estimation protocol for a homogeneous M2M network multiple times to do this is inefficient. { In this paper, we design two node cardinality estimation protocols that are specifically designed for \emph{heterogeneous} M2M networks.}

{ We consider an M2M network with $T$ types of nodes, where $T \ge 2$ is an integer, which we refer to as Type 1, Type 2, \ldots, Type $T$ nodes; e.g., these may be emergency, periodic, normal data type nodes etc.} The node cardinality estimation problem in heterogeneous M2M networks is defined, and the Lottery Frame (LoF) based protocol~\cite{qian2011cardinality}, which is a node cardinality estimation protocol for homogeneous networks, and which we extend for node cardinality estimation in heterogeneous networks in this paper, is reviewed in \Sectref{Est_Prob}. { In \Sectref{EstSchemeforT}, we design two node cardinality estimation protocols to rapidly obtain separate estimates of the number of active nodes of each traffic type in a heterogeneous M2M network, viz., Method I (see \Sectref{EstSchemeMethod1}), which is a simple scheme, and Method II (see \Sectref{EstSchemeMethod2}), which is more sophisticated and performs better than Method I.} We compute a closed form expression for the expected number of time slots required by our first estimation protocol, Method I, to execute (see Section~\ref{secEKERforT}) as well as a simple upper bound on it, which shows that the expected number of time slots required by the protocol to obtain the above estimates is small (see Section~\ref{UB_EstforT}). Next, in Section~\ref{MAC_Proto}, we use our estimation protocols as part of a Cognitive MAC protocol that we design for heterogeneous M2M networks.
In the proposed MAC protocol, time is divided into frames of equal duration, with each frame containing an estimation window (in which active node cardinalities are estimated using one of the estimation schemes proposed in \Sectref{EstSchemeforT}), a contention window (CW) and a data transmission window (DTW).  Whenever a node succeeds in contention on a given channel during the CW, the BS reserves the requested number of time slots for data transmission by that node in the DTW. Slotted ALOHA~\cite{bertsekas1992data} is used for contention in the CW, and the contention probability used by each node is the reciprocal of the estimated number of contending nodes on the channel;~\emph{thus, the estimates obtained using our estimation protocols are used for optimizing the contention probabilities}. We mathematically analyze the performance of the proposed MAC protocol and obtain expressions for the expected number of successful contentions per frame and { the expected amount of energy consumed (see Section~\ref{Ana_Res}). Using simulations, we evaluate the performances of Method I and Method II, and compare them with the performance of a protocol in which the LoF based protocol~\cite{qian2011cardinality} is separately executed $T$ times to obtain the active node cardinality of each type, in \Sectref{Simu1}. Our simulations show that both Method I and Method II significantly outperform the $T$ repetitions of LoF based protocol, in terms of execution time, whenever the probabilities that nodes are active in a given frame are sufficiently low, as would be the case in M2M networks in which the nodes are, e.g., sensors that occasionally transmit measurements or nodes that transmit alarms, emergency alerts and other infrequent messages.} Also, via simulations, we evaluate the performance, in terms of average throughput and average delay, of our MAC protocol and compare it with that of a hypothetical ``ideal protocol'', which is assumed to know the exact number of active nodes at any time (see Section~\ref{Simu2}).  Finally conclusions and directions for future research are provided in \Sectref{section:Conclusions}.

\section{Related Work}\label{ReWo}
%
A scheme to estimate the number of active nodes in an M2M network is proposed in~\cite{park2014enhancement}. In the proposed method every active device selects a slot uniformly at random from a set of slots and transmits a Power Save-poll message in the selected slot. The access point (AP) estimates the number of active nodes by using the number of empty slots and the maximum likelihood (ML) estimation method. In~\cite{wu2013fasa}, an iterative method is proposed for active node cardinality estimation in an M2M network, and is based on using drift analysis on the access results (in particular, statistics of consecutive empty and collision slots) of the past slots. { In~\cite{Est_CSMA}, a modified CSMA/CA protocol for an M2M network is proposed, wherein the size of the backoff window is computed using the size of the preceding backoff window and previous active node cardinality estimates. The size of the current backoff window is, in turn, used to estimate the current number of active nodes. In~\cite{Congestion_Liu}, a novel scheme for congestion reduction and active node cardinality estimation that uses a 6-dimensional Markov chain is proposed.} In~\cite{hsu2013adaptive}, a MAC protocol, which incorporates an active node cardinality estimation scheme, for an M2M network that uses multiple channels is proposed.
In the proposed scheme, during active node cardinality estimation, only one channel is used and all the other channels remain unused. In contrast, in the schemes proposed in this paper, all the available channels are used during active node cardinality estimation, due to which the utilization of spectrum is improved.

The problem of tag  cardinality estimation in Radio-Frequency Identification (RFID) systems is similar to that of active node cardinality estimation in M2M networks. In the former context, an RFID system reader estimates the number of RFID tags, similar to the latter context, in which a BS estimates the number of active nodes in an M2M network. Tag cardinality estimation methods for RFID systems have been proposed in~\cite{qian2011cardinality, kodialam2006fast, kodialam2007anonymous, li2010energy, zheng2012pet,  zhou2016understanding, RFID_Block_Liu, RFID_Counterfeit_Gong}. 

However, all the node (or tag) cardinality estimation schemes studied in the above papers are for a \emph{homogeneous} network, wherein all nodes (or tags) are alike. In contrast, in this paper, we propose two estimation schemes for a \emph{heterogeneous} network with $T$ different types of nodes, which efficiently compute separate estimates of the number of active nodes of each type.

Extensive surveys on MAC protocol design for M2M networks are provided in~\cite{rajandekar2015survey, islam2014survey, Survey_Verma, Survey_Wang}. Also, several mechanisms for achieving improved M2M communications in LTE and LTE-Advanced cellular networks have been proposed; see~\cite{hasan2013random} for a survey. { In~\cite{SDMAC_Liu}, an adaptive database-driven MAC protocol for cognitive M2M networks is proposed. Nodes adaptively use either the spectrum database or the local sensing approach for obtaining the channel availability information in order to maximize the network throughput. An energy efficient and load adaptive MAC protocol for cellular-based M2M networks is proposed in~\cite{E2Mac_Miao}. Nodes are divided into clusters; for intra-cluster communications, a multi-phase CSMA/CA protocol proposed in the paper is used, whereas for inter-cluster communications a conventional cellular access scheme is used.} In~\cite{azquez2013dpcf}, a hybrid MAC protocol that uses contention-based channel access (CSMA/CA) when the network load is low and reservation-based access when the load is high is proposed. In~\cite{liu2013scalable}, a hybrid MAC protocol, in which each time frame consists of a contention period followed by a transmission period, is proposed. The devices that successfully contend in the contention period are assigned a time slot for data transmission in the transmission period. { A similar hybrid MAC protocol is proposed in~\cite{Hybdrid_Saad}, wherein slotted ALOHA is used in the contention period and reservation-based time division multiple access (TDMA) is used in the transmission period.} The protocol proposed in~\cite{liu2013scalable} is extended to heterogeneous M2M networks in~\cite{liu2014design}, wherein different types of devices with different service requirements and priorities are considered. In~\cite{so2004multi}, a MAC protocol is proposed for multichannel ad hoc networks and this is modified in~\cite{hsu2013adaptive} for use in M2M networks. In the protocol proposed in~\cite{hsu2013adaptive}, time is divided into frames and they are further divided into three phases, viz., estimation phase, contention phase and data transmission phase. The number of active users is  estimated in the estimation phase. In the contention phase, all the active users tune to a common control channel and contend for channel access using contention probabilities which are obtained as a function of the number of estimated nodes. The nodes which are successful in contention transmit their data packets in the data transmission phase in parallel on different channels. In the protocol proposed in~\cite{wang2010adaptive}, time is divided into slots, and in each slot, nodes contend using a contention probability, which is based on a statistical estimate of the present traffic load, and then transmit a Request to Send (RTS) packet, which is responded to with a Clear to Send (CTS) packet, and these packets are followed by transmission of a data packet.
In~\cite{park2014enhancement}, the 802.11ah MAC protocol is modified for M2M communications as follows: first, estimation of the number of active M2M devices is done and then this estimate is used to adapt the length of the Restricted Access Window, in which only M2M devices are allowed to contend. 
In~\cite{wu2013fasa}, a modified version of the Slotted-ALOHA scheme is presented, in which results of the previous slots are considered to estimate the transmission attempt probability that maximizes the throughput in the current slot. 
In~\cite{lo2011enhanced}, an overload control mechanism is presented for M2M communication in LTE-Advanced networks, in which based on the traffic load on the random access channel (RACH), the base station adjusts the number of RACH resources.

However, to the best of our knowledge, our Cognitive MAC protocol is the first to employ separate estimates of the numbers of active nodes of different types for selecting the optimal contention probabilities in a heterogeneous M2M network.


\section{Problem Formulation and Background}\label{Est_Prob}
The estimation problem in heterogeneous M2M networks is defined in Section~\ref{nwmodel}. In Section~\ref{SubSec_Est}, the Lottery Frame (LoF) based protocol~\cite{qian2011cardinality},~\cite{flajolet1985probabilistic}, which is a node cardinality estimation scheme for homogeneous networks, and which we extend to estimate node cardinalities in heterogeneous M2M networks, is briefly described. 

\begin{figure}[tbp]
  \centering
    \includegraphics[width=0.3\textwidth]{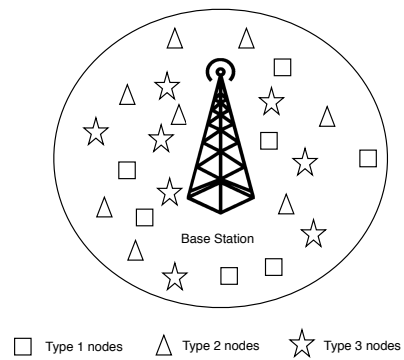}
    \caption{The figure shows a base station and $T = 3$ types of nodes within its range (the area inside the circle).}
    \label{Network_Model} 
\end{figure}

{ \subsection{The Estimation Problem in Heterogeneous M2M Networks}\label{nwmodel}}
{ Consider a heterogeneous M2M network consisting of a base station (BS) and $T$ different types, say Type 1, Type 2, \ldots, Type $T$, of M2M devices (nodes) in its range as shown in \Figref{Network_Model} for the case $T = 3$. We denote the sets of nodes of Type 1, Type 2, \ldots, Type $T$ as $\mathcal{N}_1$, $\mathcal{N}_2$, \ldots, $\mathcal{N}_T$ respectively; let~\footnote{$|A|$ denotes the cardinality of set $A$.} $|\mathcal{N}_b| = N_b$, $b \in \{1,2, \ldots, T\}$. 
Time is divided into frames of equal duration, and in each frame only a subset of the nodes of each type are \emph{active}, i.e., have data to send to the BS. Also, each frame is divided into time slots of equal durations. Let $n_b$ be the number of active nodes of Type $b$, $b \in \{1,2, \ldots, T\}$, in a given frame. Our objective is to design estimation protocols to estimate the values of $n_1$, $n_2$,  \ldots,  $n_T$ rapidly, i.e., using a small number of time slots.} 

\subsection{Review of the LoF Based Protocol}\label{SubSec_Est}
The LoF based estimation protocol was designed in~\cite{qian2011cardinality} and uses the probabilistic bitmap counting technique proposed in~\cite{flajolet1985probabilistic} for tag cardinality estimation in  RFID systems. The LoF based estimation protocol is designed for a homogeneous network. Our proposed protocols extend the LoF based protocol to a heterogeneous network with $T$ types of nodes. So we provide a brief review of the LoF based protocol in this subsection. 

Every tag (equivalent to a node in M2M networks) has a unique binary identification (ID) number that is $l$ bits in length. The hash value $h$ of any tag is defined as the `position of the least significant zero bit' in its ID. For example, $h(01001001) = 1$ and $h(00101111) = 4$, where $h(I)$ denotes the hash value corresponding to ID $I$. So if $h$ is the hash value of a random tag, then assuming that each of the $l$ bits of the corresponding ID independently equals $0$ or $1$ with probability ${1}/{2}$ each,  $P(h=i) = {1}/{2^{(i+1)}}$, $i = 0, 1, 2, \ldots, l-1$  and~\footnote{If all the bits of the ID are $1$, then its hash value is defined to be $l$.} $P(h=l) ={1}/{2^l}$. 

Now, time is divided into slots of equal durations. During the estimation process, each active tag with hash value $h$ transmits a packet in the $h^{th}$ time slot, for $h = 0, 1, 2, \ldots, l$. A corresponding bitmap ($BM$) of $0$s and $1$s is generated by the RFID system reader (equivalent to the BS in M2M networks) based on the slot results; the $h^{th}$ bit of the $BM$ is  $0$ if the $h^{th}$ time slot is empty (i.e., one in which no tag transmits) and $1$ if the $h^{th}$ time slot is non-empty (i.e., one in which one or more tags transmit). Let $\rho = \min\{h|BM(h) = 0\}$, where $BM(h)$ is the $h^{th}$ bit of the above bitmap; then the estimated value of $n$ (the actual number of active tags) is $\hat{n} = 1.2897 \times 2^{\rho}$~\cite{qian2011cardinality}.  It is also proved in~\cite{qian2011cardinality} that the LoF based protocol executes within $\ceil{\log_2 {n_{all}}}$~\footnote{$\ceil{x} =$ The smallest integer greater than or equal to $x$.} slots, where $n_{all}$ is the total number of all possible binary IDs. 

{ Note that if the LoF based protocol is executed $T$ times to separately estimate $n_1$, $n_2$, \ldots, $n_T$ in the network model for M2M networks described in \Sectref{nwmodel}, then $\sum_{b=1}^{T} \ceil{\log_2 ({n_{b,all}})}$ slots are required, where $n_{b,all}$ is the total number of all possible binary IDs of the $b^{th}$ type of nodes. To reduce the number of slots, we propose two fast node cardinality estimation schemes, which are described in the following  section.}

\section{Fast Node Cardinality Estimation Schemes}\label{EstSchemeforT}
In this section, we present two fast node cardinality estimation schemes for heterogeneous M2M networks, viz., Method I and Method II which are described in Sections~\ref{EstSchemeMethod1} and~\ref{EstSchemeMethod2} respectively. Method I is a simple scheme and Method II is more sophisticated and performs better than Method I. A closed form expression for the expected number of time slots required by our first estimation protocol, Method I, to execute is computed in \Sectref{secEKERforT} and a simple upper bound on it is established in \Sectref{UB_EstforT}. 

{ \subsection{Method I}\label{EstSchemeMethod1}}
{ We now describe the scheme used in Method I. Let $T \ge 2$ be arbitrary. The estimation process is carried out in three phases, which we describe in Sections~\ref{FirstPhaseforT},~\ref{SecondPhaseforT} and~\ref{ThirdPhaseforT}.  We refer to the set of slots used during the estimation process as the Estimation Window (EW). The structure of a typical EW is shown in Fig.~\ref{EST_Window} for the case $T = 3$.
At the end of the estimation process, separate estimates, say $\hat{n}_1, \hat{n}_2$, $\ldots$, $\hat{n}_T$, of the number of active nodes of the $T$ types, $n_1, n_2$, $\ldots$, $n_T$ (see Section~\ref{nwmodel}), are obtained. For each $b \in \{1, 2, \ldots T\}$, the estimate $\hat{n}_b$ \emph{is equal to (and hence, as accurate as)} the estimate of $n_b$ that would have been obtained if the LoF protocol~\cite{qian2011cardinality},~\cite{flajolet1985probabilistic} were used for the estimation. However, note that under mild conditions, the total number of time slots used in Method I is much smaller than the number of time slots that would have been required if the LoF protocol were separately executed $T$ times to estimate $n_1, n_2$, $\ldots$, $n_T$.
\begin{figure}[!t]
\centering
\resizebox{0.95\columnwidth}{!}{\includegraphics{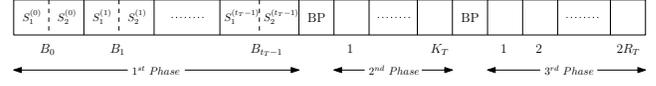}}
\caption{The figure shows the structure of the Estimation Window for $T = 3$. ``BP'' denotes a Broadcast Packet. Note that in the $1^{st}$ phase, each block consists of $T - 1 = 3-1 = 2$ slots.}
    \label{EST_Window} 
\end{figure}

At a high level, Method I operates as follows. Let $t_T = \ceil{\log_2 (\max (n_{1, all}, n_{2, all}, \ldots, n_{T, all}))}$. Also, for $b \in \{1, 2, \ldots T\}$ and $i \in \{0, 1, \ldots, t_T-1\}$, let $B(b,i)$ be $1$ (respectively, $0$) if the $i^{th}$ slot would have been non-empty (respectively, empty) if the LoF protocol were used to estimate the number of active nodes of Type $b$. From Section~\ref{SubSec_Est}, it is clear that if the bit patterns $B(b,i)$, $b \in \{1, 2, \ldots T\}$, $i \in \{0, 1, \ldots , t_T-1\}$, are known, then the LoF estimates $\hat{n}_1, \hat{n}_2$, $\ldots$, $\hat{n}_T$, of $n_1, n_2, \ldots,  n_T$ respectively, can be deduced. In our estimation scheme, the bit patterns $B(b,i)$, $b \in \{1, 2, \ldots T\}$, for most values of $i$ are found in the first phase; ambiguity about the rest remains, which is resolved in the second and third phases.    

\subsubsection{First Phase}\label{FirstPhaseforT}
In the first phase, $(T-1) t_T$ slots are used. These $(T-1) t_T$ slots are divided into $t_T$ disjoint sets of $(T-1)$ consecutive slots, each called a ``block'' (see Fig.~\ref{EST_Window} for $T=3$); let $B_i$ denote the $i^{th}$ block. For each $i \in \{0, 1, \ldots, t_T-1\}$, active nodes from $\mathcal{N}_1$ whose hash value is $i$ send a packet containing the symbol $\alpha$ in each of the $T-1$ slots of $B_i$. Also,  active nodes from $\mathcal{N}_2$, (respectively, $\mathcal{N}_3$, $\ldots$, $\mathcal{N}_T$)  whose hash value is $i$ send a packet containing the symbol $\beta$ only in the first slot (respectively, only in the second slot, \ldots, only in the $(T-1)^{th}$ slot) of $B_i$. So every slot has four possible outcomes, which are as follows:  (i) Empty ($E$) if no node transmits in the slot, (ii) Collision ($C$) if  two or more nodes transmit, (iii) $\alpha$ if exactly one node of Type 1 transmits, (iv) $\beta$  if exactly one node of Type 2 or Type 3 or $\ldots$ Type $T$ transmits. 
The possible outcomes in a block for $T=3$ are shown in the first two columns of \Tabref{Tab_BitPatterns}~\footnote{Note that the block results ($E$, $\alpha$), ($\alpha$, $E$), ($\alpha$, $\beta$) and ($\beta$, $\alpha$) cannot occur under the above protocol.}. Note that for $b \in \{1, 2, \ldots T\}$, $i \in \{0, 1, \ldots, t_T-1\}$, $B(b,i)$ equals $1$ (respectively, $0$) if and only if at least one node (respectively, no node) of Type $b$ transmits in block $B_i$. The bit patterns $B(b,i)$, $b \in \{1, 2, \ldots T\}$ corresponding to each possible block outcome for $T=3$ are shown in the last three columns of \Tabref{Tab_BitPatterns}.
For example if Slot1 results in $C$ and Slot2 results in $\alpha$, then it implies that exactly one node from $\mathcal{N}_1$, at least one node from $\mathcal{N}_2$ and none from $\mathcal{N}_3$ have transmitted. Similarly if both the slots result in $\beta$, then it implies that exactly one node each from $\mathcal{N}_2$ and $\mathcal{N}_3$, and none from $\mathcal{N}_1$ have transmitted. 

\begin {table} [ht]
\centering
\caption{$E$, $C$ and $\star$ denote ``Empty'', ``Collision'' and  ``ambiguous result'' respectively.}
\begin{tabular}{|c|c|c|c|c|} 
\hline
\multicolumn{2} { | c | } {Outcome in block $B_i$}  & \multicolumn{3} { | c |} {Bit patterns}\\ 
\hline
Slot1 & Slot2 & $B(1,i)$ & $B(2,i)$ & $B(3,i)$ \\
\hline
$E$ & $E$ & 0 & 0 & 0 \\
\hline
$E$ & $C$ & 0 & 0 & 1 \\
\hline
$E$ & $\beta$ & 0 & 0 & 1 \\
\hline
$C$ & $E$ & 0 & 1 & 0 \\
\hline
$C$ & $C$ & $\star$ & $\star$ & $\star$ \\
\hline
$C$ & $\alpha$ & 1 & 1 & 0 \\
\hline
$C$ & $\beta$ & 0 & 1 & 1 \\
\hline
$\alpha$ & $C$ & 1 & 0 & 1 \\
\hline
$\alpha$ & $\alpha$ & 1 & 0 & 0 \\
\hline
$\beta$ & $E$ & 0 & 1 & 0 \\
\hline
$\beta$ & $C$ & 0 & 1 & 1 \\
\hline
$\beta$ & $\beta$ & 0 & 1 & 1 \\
\hline
\end{tabular}
\label{Tab_BitPatterns}
\end{table}

The outcome ($C$, $C$, \ldots, $C$) in which collisions occur in all $T-1$ slots of a block may be due to transmissions by at least two nodes from $\mathcal{N}_1$, or by one node from $\mathcal{N}_1$ and at least one node each from $\mathcal{N}_2$, $\mathcal{N}_3$, \ldots, $\mathcal{N}_T$, or by at least two nodes each from $\mathcal{N}_2$, $\mathcal{N}_3$, \ldots, $\mathcal{N}_T$ and none from $\mathcal{N}_1$. Due to this ambiguity, if the outcome ($C$, $C$, \ldots, $C$) occurs in the block $B_i$, then the second phase is used to find the bit patterns $B(1, i)$, $B(2, i)$, \ldots, $B(T, i)$.  Note that \Tabref{Tab_BitPatterns} provides the bit patterns $B(1,  i)$, $B(2, i)$ and $B(3, i)$ for every possible outcome, except the outcome ($C, C$) in block $B_i$ for the case $T=3$; similarly, it is easy to check that for the case where $T$ is arbitrary, the bit patterns $B(1, i)$, $B(2, i)$, \ldots, $B(T, i)$ for every possible outcome, except the outcome ($C, C, \ldots, C$), in block $B_i$ can be unambiguously found. Let $C_{I}$ be the set of block numbers $i$ in which the ambiguous outcome ($C$, $C$, \ldots, $C$) has occurred. A broadcast packet (BP) of length $\ceil{t_T/S_W}$ slots, where $S_W$ is the number of bits that can be transmitted in a time  slot~\footnote{{ For example, $S_W$ may be $5$~\cite{kodialam2007anonymous}.}},    is sent by the BS after the first phase (see \Figref{EST_Window}), which contains a bitmap of length $t_T$ in which the $i^{th}$ bit is $1$ (respectively, $0$) if $i \in C_I$ (respectively, $i \notin C_I$).  

\subsubsection{Second Phase}\label{SecondPhaseforT}
In the second phase, only the active nodes from $\mathcal{N}_1$ whose hash value belongs to the set $C_{I}$ participate. Specifically, for each $j = 1, 2, \dots , |C_I|$, in the $j^{th}$ slot of the second phase, the active nodes from $\mathcal{N}_1$ whose hash value equals the block number of the $j^{th}$ block whose outcome was ($C$, $C$, $\ldots$, $C$) in the first phase transmit. Nodes from $\mathcal{N}_2$, $\mathcal{N}_3$, $\ldots$, $\mathcal{N}_T$ do not transmit in the second phase.

Now, consider the slot in the second phase corresponding to the  block $B_i$ in the first phase, where $i \in C_I$. If the slot result is empty, then it follows that $B(1,i) =  0$, $B(2,i) = 1$, $\ldots$, $B(T,i) = 1$; also, if the slot result is one packet transmission, then $B(1,i) =  1$, $B(2,i) = 1$, $\ldots$, $B(T,i) = 1$. If the slot result is $C$, then $B(1,i) =  1$; however,  ambiguity about the values of $B(2,i)$, $\ldots$, $B(T,i)$ still remains and it is resolved in the third phase. 
Let $C_{II} \subseteq C_I$ be the set of block numbers $i$ for which a collision occurred in the second phase. A BP  of length $\ceil{K_T/S_W}$ slots is sent by the BS after the second phase (see \Figref{EST_Window}), which contains a bitmap of length $|C_I|$ in which the $j^{th}$ bit is $1$ (respectively, $0$) if the result of the $j^{th}$ slot of the second phase was (respectively, was not) a collision.  

\subsubsection{Third Phase}\label{ThirdPhaseforT}
In this phase, only those active nodes from $\mathcal{N}_2$, $\mathcal{N}_3$, $\ldots$, $\mathcal{N}_T$ participate, for which the result of the corresponding block in the first phase was ($C$, $C$, \ldots, $C$) and the result of the  corresponding slot in the second phase was a collision. That is, the active nodes from $\mathcal{N}_2$, $\mathcal{N}_3$, $\ldots$, $\mathcal{N}_T$  whose hash value belongs to $C_{II}$ participate. For ease of understanding, we first describe the procedure in the case $T=3$ and then generalize to the case where $T$ is arbitrary. 
In the case $T=3$, the odd (respectively, even) numbered slots of the third phase are used by nodes from $\mathcal{N}_2$ (respectively, $\mathcal{N}_3$). Specifically, for each  $j = 1, 2, \dots , |C_{II}|$, in slot $2j -1$ (respectively, $2j$) of the third phase, the active nodes from $\mathcal{N}_2$ (respectively, $\mathcal{N}_3$) whose hash value equals the first phase block number, say $i$,  of the $j^{th}$ element of $C_{II}$ transmit. If slot $2j -1$ is empty, then $B(2,i) = 0$, else $B(2,i) = 1$. Similarly, if slot  $2j$ is empty, then $B(3,i) = 0$, else $B(3,i) = 1$. In the case where $T$ is arbitrary, for each  $j \in \{1, 2, \dots , |C_{II}|\}$, the $((j-1)(T-1)+1)^{th}$, $((j-1)(T-1)+2)^{th}$, \ldots $(j(T-1))^{th}$ slots of the third phase are used by nodes from $\mathcal{N}_2$, $\mathcal{N}_3$, $\ldots$, $\mathcal{N}_T$ respectively.
Specifically, for each  $j = 1, 2, \dots , |C_{II}|$, in slot $((j-1)(T-1)+1)$ (respectively, $((j-1)(T-1)+2)$, \ldots, $(j(T-1)$)) of the third phase, the active nodes from $\mathcal{N}_2$ (respectively, $\mathcal{N}_3$, \ldots, $\mathcal{N}_T$) whose hash value equals the first phase block number, say $i$,  of the $j^{th}$ element of $C_{II}$ transmit. If slot $((j-1)(T-1)+1)$ (respectively, $((j-1)(T-1)+2)$, \ldots, $(j(T-1)$)) is empty, then $B(2,i) = 0$ (respectively, $B(3,i) = 0$, \ldots, $B(T,i) = 0$), else $B(2,i) = 1$ (respectively, $B(3, i) = 1$, \ldots, $B(T, i) = 1$). Also, since $B(1,i) = 1$, the above ambiguity is resolved in the third phase.

\subsubsection{Determination of Expected Number of Time Slots Required by Method I to Execute}\label{secEKERforT}
Recall that ($T-1$)$t_T$ slots are required in the first phase. Let $K_T$ (respectively, ($T-1$)$R_T$) be the number of slots required in the second phase (respectively,  third phase). For tractability, in this subsection and in \Sectref{UB_EstforT}, we assume that $n_{1, all} = n_{2, all} = \ldots = n_{T, all}$.

\paragraph{Determination of $E[K_T]$}
Note that $0 \le K_T \le t_T$.
Let $S_1^{(i)}, S_2^{(i)}, \ldots, S_{T-1}^{(i)}$ represent the result of the first, second, $\ldots$ , $(T-1)^{th}$ slot of $B_i$ respectively. Also, let $I_{\mathcal{F}}$ denote the indicator random variable corresponding to event $\mathcal{F}$, i.e., 
\[
I_{\mathcal{F}}= \left\{ \begin{array}{ll}
1, & \mbox{if } \mathcal{F} \mbox{ occurs}, \\
0, & \mbox{else.} \\
\end{array}
\right.\]
Clearly, $K_T = \sum_{i=0}^{t_T-1} I_{\{S_1^{(i)} = C, S_2^{(i)} = C, \ldots , S_{T-1}^{(i)} = C\}}$. So:
\begin{equation}
\label{EKforT}
E[K_T] = \sum_{i=0}^{t_T-1} P(S_1^{(i)} = C, S_2^{(i)} = C, \ldots , S_{T-1}^{(i)} = C).
\end{equation}
The conditions under which collisions occur in all the ($T-1$) slots of $B_i$ are as follows: $1)$ At least two nodes from $\mathcal{N}_1$ transmit in $B_i$, $2)$
Exactly one node from $ \mathcal{N}_1$ and at least one node each from $\mathcal{N}_2, \mathcal{N}_3, \ldots , \mathcal{N}_{T}$ transmit in $B_i$, $3)$ At least two nodes each from  $\mathcal{N}_2, \mathcal{N}_3, \ldots , \mathcal{N}_{T}$ and none from  $\mathcal{N}_1$ transmit in $B_i$. Let $Q_{1}(i)$, $Q_{2}(i)$ and $Q_{3}(i)$ denote the probabilities of the events in $1)$, $2)$ and $3)$ respectively. Then: 
\begin{equation}
\label{eqPforT}
P(S_1^{(i)} = C, S_2^{(i)} = C, \ldots , S_{T-1}^{(i)} = C) =  Q_{1} (i) + Q_{2} (i) + Q_{3} (i).
\end{equation}
It is easy to show that $Q_{1}(i) = 1 - u{(n_1 ,i)} - v{(n_1, i)}$, $Q_{2} (i) = v{(n_1, i)} (1 -  u{(n_2, i)}) (1 -  u{(n_3, i)}) \ldots (1 -  u{(n_T, i)}) $ and $Q_{3} (i) = u{(n_1, i)}(1 - u{(n_2, i)} - v{(n_2, i)})  (1 - u{(n_3, i)} - v{(n_3, i)}) \ldots (1 - u{(n_T, i)} - v{(n_T, i)})$, where $u{(n, i)} =  \big(1 - {p_i}\big)^{n}$, $v{(n, i)} = n p_i\big(1 - p_i\big)^{{n} -1}$, and 
\begin{equation}
\label{EQ:pi}
p_i = \left\{ \begin{array}{ll}
{1}/{2^{(i+1)}}, & \mbox{for } i = 0, 1, 2, \ldots, t_T - 2,\\
{1}/{2^{t_T - 1}} & \mbox{for } i = t_T - 1, \\
\end{array}
\right.
\end{equation}
is the probability that the hash value of a random node is $i$ (see \Sectref{SubSec_Est}). By \eqref{EKforT} and \eqref{eqPforT}, 
the expected number of slots required in the second phase is:
\begin{equation}
\label{eqEKforT}
E[K_T] = \sum_{i=0}^{t_T-1}  \{Q_{1} (i) + Q_{2} (i) + Q_{3} (i)\}.
\end{equation}

\paragraph{Determination of $E[R_T]$}
 Note that $0 \le R_T \le K_T$. It is easy to show that: 
\begin{equation}\label{eqERforT}
E[R_T] = \sum_{i=0}^{t_T-1}  Q_{1} (i).
\end{equation}
The expected total number of slots required by Method I to execute is $(T-1)t_T  + \ceil{t_T/S_W} + E[K_T] +  \ceil{E[K_T]/S_W} + (T-1)E[R_T]$ (see Fig.~\ref{EST_Window}), where $E[K_T]$ and $E[R_T]$ are given by \eqref{eqEKforT} and \eqref{eqERforT} respectively.


\subsubsection{Upper Bound on Expected Number of Time Slots Required by Method I to Execute} \label{UB_EstforT}
Although the expressions derived in \Sectref{secEKERforT} are exact, they are complicated. So to gain insight, in this subsection, we provide simple upper bounds on $E[K_T]$ and $E[R_T]$ and use them to obtain an upper bound on the expected total number of slots required by Method I to execute. Let $n_r = \max{(n_1, n_2, \ldots , n_T)}$,  $l_y = \ceil{(\log_2y)}$, and $s = t_T - l_{n_r}$.
\begin{theorem}
\label{ApproxEKforT}
$E[K_T]  \le l_{n_r} - 1+  \frac{2}{3} \frac{n_1^2}{n_r^2}\Big(1+2/4^s\Big) + \Big(\frac{n_2 n_3 \ldots n_T}{n_r^{T-1}}\Big)^2 \frac{1}{2^{T-1} (1 - 4^{-(T-1)})} \bigg(1 - \frac{2}{4^{(T-1)s}} \Big(1 - \frac{4^{T-1}}{2}\Big)\bigg) + \Big(\frac{n_1 n_2 \ldots n_T}{n_r^T}\Big)  \frac{1}{1 - 2^{-T}} \Big( 1 - \frac{2}{2^{Ts}} (1 - 2^{T-1})\Big)$.
\end{theorem}
\begin{theorem}
\label{ApproxERforT}
$E[R_T] \le l_{n_r} - 1 +  \frac{2}{3} \frac{n_1^2}{n_r^2}\Big(1+2/4^s\Big)$.
\end{theorem}
The proofs of Theorems~\ref{ApproxEKforT} and~\ref{ApproxERforT} are provided in the Appendix. Note that an upper bound on the expected total number of slots required by Method I to execute, which is  $(T-1)t_T  + \ceil{t_T/S_W} + E[K_T] +  \ceil{E[K_T]/S_W} + (T-1)E[R_T]$, follows from Theorems~\ref{ApproxEKforT} and~\ref{ApproxERforT}. In Section~\ref{Simu1}, using numerical computations, we study as to how tight this upper bound is.

\subsection{Method II}\label{EstSchemeMethod2}
\begin{figure}[!t]
\centering
\resizebox{1\columnwidth}{!}{\includegraphics{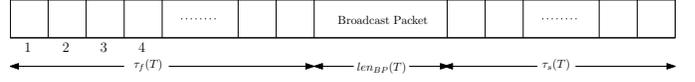}}
\caption{{ The figure shows the structure of the Estimation Window used in Method II for $T \geq 4$.}}
    \label{EST_Window_2} 
\end{figure}

\begin{figure}[!t]
\centering
\resizebox{1\columnwidth}{!}{\includegraphics{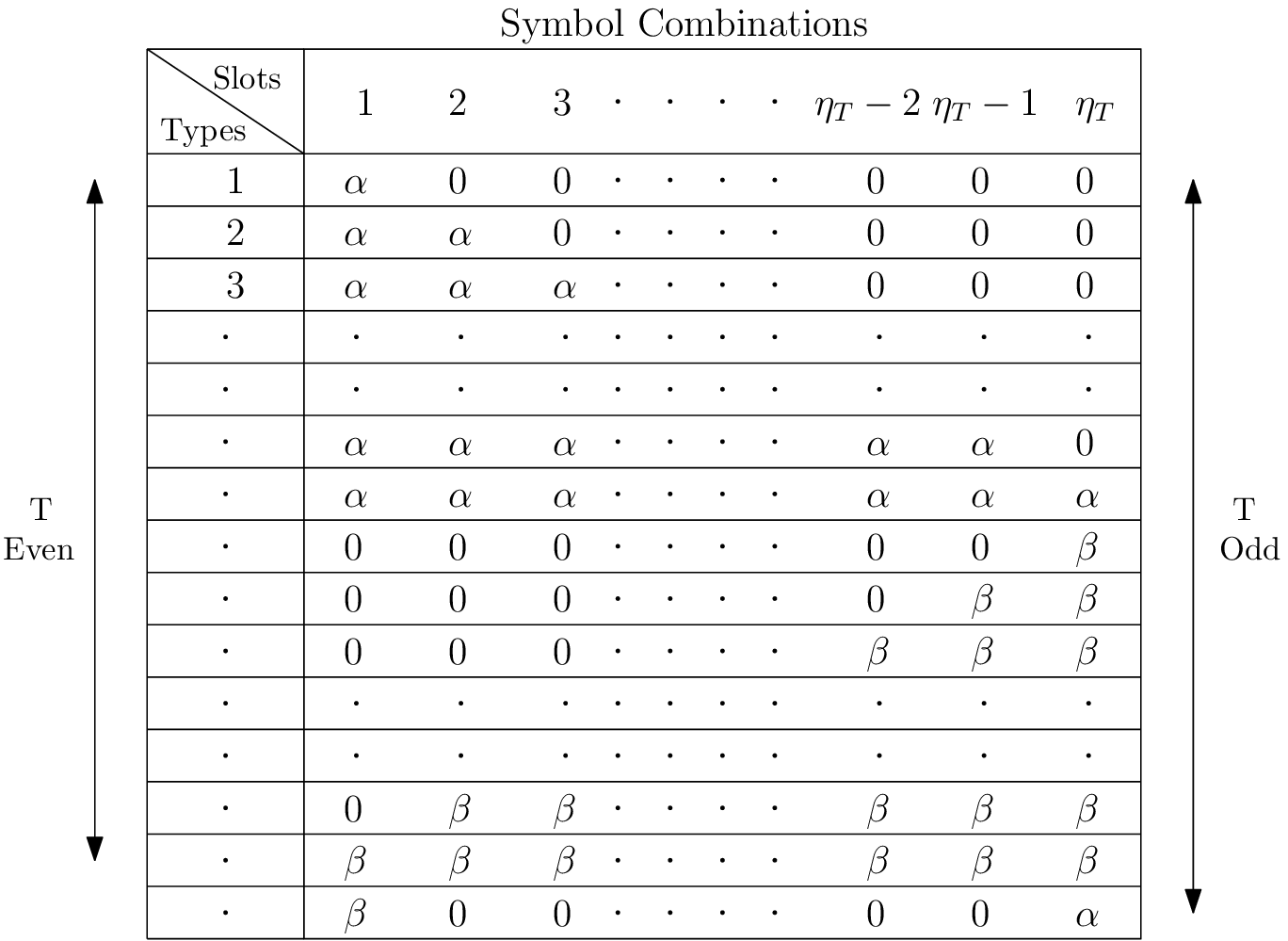}}
\caption{{ The figure shows the symbol combinations used by each type in Method II. $\eta_T = T/2$ if $T$ is even and $\eta_T = (T-1)/2$ if $T$ is odd. The symbol $0$ indicates ``no transmission''.}}
    \label{Sym_Combo} 
\end{figure}

The estimation scheme used for $T = 2$ and $T = 3$ in Method II is the same as that used in Method I. We now describe the scheme used in Method II for $T \ge 4$. For ease of understanding, we provide the estimation schemes used under Method II for $T = 4, 5$ and $6$ in Sections~\ref{Method2_4},~\ref{Method2_5} and~\ref{Method2_6} respectively in detail. The estimation schemes used for $T \ge 7$ are similar. Similar to Method I, at the end of the estimation process of Method II, separate estimates, say $\hat{n}_1, \hat{n}_2$, $\ldots$, $\hat{n}_T$, of the number of active nodes of the $T$ types, $n_1, n_2$, $\ldots$, $n_T$ (see Section~\ref{nwmodel}), are obtained. For each $b \in \{1, 2, \ldots T\}$, the estimate $\hat{n}_b$ is equal to (and hence, as accurate as) the estimate of $n_b$ that would have been obtained if the LoF protocol~\cite{qian2011cardinality},~\cite{flajolet1985probabilistic} were used for the estimation. However, note that under mild conditions, the total number of time slots used in Method II is much smaller than the number of time slots that would have been required if the LoF protocol were separately executed $T$ times to estimate $n_1, n_2$, $\ldots$, $n_T$.

At a high level, Method II operates as follows. The structure of a typical EW when Method II is used for estimating node cardinalities for $T \ge 4$ is shown in Fig.~\ref{EST_Window_2}.  Method II has two phases. In the first phase $t_T$ blocks, each consisting of $\eta_T$ slots, where $\eta_T = T/2$ if $T$ is even and $\eta_T = (T-1)/2$ if $T$ is odd, are used. Hence, the first phase consists of $\tau_f(T) = (T\times t_T)/2$ slots if $T$ is even and $\tau_f(T) = ((T-1)\times t_T)/2$ slots if $T$ is odd. The number of slots in the second phase ($\tau_s(T)$) depends on the slot results of the first phase.   In each block $B_i$ of the first phase, where $i \in \{0,1,  \ldots, t_T - 1\}$, active nodes of all $T$ types with hash value $i$ transmit their corresponding symbol combinations. The symbol combinations used by each type are shown in Fig.~\ref{Sym_Combo}. For example, each active node of Type 1 and hash value $i$ transmits symbol $\alpha$ in slot 1 and does not transmit in slots $2, \ldots, \eta_T$ of block $B_i$. Each active node of Type 2 and hash value $i$ transmits symbol $\alpha$ in slots 1 and 2 and does not transmit in slots $3, \ldots, \eta_T$ of block $B_i$. For $T$ even, each active node of Type $T$ and hash value $i$ transmits symbol $\beta$ in slots $1, \ldots, \eta_T$ of block $B_i$. For $T$ odd, each active node of Type $T$ and hash value $i$ transmits symbol $\beta$ in slot 1, symbol $\alpha$ in slot $\eta_T$, and does not transmit in slots $2, \ldots, \eta_T -1$ of block $B_i$. In each slot there are four possible outcomes: 0 (no transmission), $\alpha$, $\beta$, and $C$ (collision). The BS records the outcome of each slot of each block. Let the bit patterns $B(b,i)$, $b \in \{ 1, 2, \ldots, T\}$, $i \in \{0, 1, \ldots, t_T-1 \}$, be defined as in \Sectref{EstSchemeMethod1}. Under Method II, the bit patterns $B(b,i)$ for most values of $b$ and $i$ are found in the first phase; ambiguity about the rest remains, which is resolved in the second phase. In particular, it is easy to check that if no collision occurs in any of the slots of a block $B_i$, then the bit patterns $B(b,i)$, $b \in \{ 1, 2, \ldots, T\}$, can be unambiguously determined. If a collision occurs in one or more slots of a block $B_i$, then ambiguity about $B(b,i)$ for some values of $b$ remains, which is resolved in the second phase. The active nodes corresponding to those blocks $B_i$ and types $b$ for which ambiguity regarding $B(b,i)$ remains after the first phase participate in the second phase; the others do not. At the end of the first phase, the BS sends a broadcast packet (BP) which informs the active nodes, as to the block numbers $i$ and type numbers $b$ whose corresponding nodes need to participate in the second phase.

\subsubsection{Scheme for $T = 4$}\label{Method2_4}
The first phase consists of $t_4$ blocks, each consisting of two slots. From Fig.~\ref{Sym_Combo}, it follows that the symbol combinations given in the following table are used. 
\begin{center}
  \begin{tabular}{| c | c |}
    \hline
    Type & Symbol Combination \\ \hline
    1 & $\alpha$ $0$ \\ \hline
    2 & $\alpha$ $\alpha$ \\ \hline
    3 & $0$ $\beta$ \\ \hline
    4 & $\beta$ $\beta$ \\ \hline    
  \end{tabular}
\end{center}

\paragraph{First Phase} \label{FirstPhase_4}
In each slot of the first phase there are four possible outcomes: 0, $\alpha$, $\beta$, and $C$. Hence, the total number of possible outcomes in the two slots of a block is $4^2 = 16$. It is easy to see that the bit patterns $B(b, i), b \in \{1, 2, 3, 4\}$, can be unambiguously found by the BS if no collision occurs in either of the slots of block $B_i$. So next, we consider the cases in which at least one collision occurs per block.
The number of cases in which a collision occurs in exactly one slot of a  block  is $\binom{2}{1}\times3 = 6$.  The possible outcomes in a block in these 6 cases are shown in the first two columns of  \Tabref{Tab4_OneC}. The types of nodes listed under the `Active' (respectively, `Inactive') column are unambiguously identified by the BS as being active (respectively,  inactive) at the end of the first phase; the types listed under the `Not Sure' column are those for which ambiguity exists, and this ambiguity needs to be resolved in the second phase. 
For example, if Slot 1 results in $C$ and Slot 2 results in $\alpha$, then using the above table of symbol combinations, the BS infers that at least one node of Type 1 and exactly one node of Type 2 are active, and all Type 3 and Type 4 nodes are inactive. Similarly, if Slot 1 results in $C$ and Slot 2 results in $\beta$, then the BS infers that at least one node of Type 1 is active, no node of Type 2 is active, and exactly  one node of Type 3 or Type 4 is active.

\begin {table} [h]
\centering
\caption{{ This table shows the inferences drawn by the BS regarding the activity or inactivity of various types of nodes  for each of the outcomes in which  exactly one collision occurs per block.}}
\begin{tabular}{|c|c|c|c|c|} 
\hline
\multicolumn{2}{|c|}{Outcome in Block $B_i$} & \multicolumn{3}{|c|}{Types} \\ \hline
Slot 1               & Slot 2              & Active & Inactive    & Not Sure         \\ \hline
C                    & 0                   & 1    & 2, 3, 4    & -                \\ \hline
C                    & $\alpha$            & 1, 2  & 3, 4   & -                \\ \hline
C                    & $\beta$             & 1   & 2    & One of \{3, 4\}   \\ \hline
0                    & C                   & 3    & 1, 2, 4   & -                \\ \hline
$\alpha$             & C                   & 3    & 4   & One of \{1, 2\}     \\ \hline
$\beta$              & C                   & 3, 4  &1, 2   & -                \\ \hline
\end{tabular}

\label{Tab4_OneC}
\end{table}

\Tabref{Tab4_TwoC_Combo} shows the possible combinations of numbers of active nodes of different types for which the result  $C C$ may occur in a block of the first phase, e.g., at least two nodes of Type 2 (second row) or exactly one node each of Types 2 and 4 (fourth row) or at least two nodes of each of Types 1 and 3 and none of Types 2 and 4 (last row). From this table, it can be seen that if the result $C C$ occurs in a block, then ambiguity remains about the activity or inactivity of all four node types, and this ambiguity needs to be resolved in the second phase.  

\begin{table}[h!]
\centering
\caption{{ This table shows the possible combinations of numbers of active nodes of different types for which the block result  $C C$ may occur. $\star$ denotes any number of active nodes.}}
\begin{tabular}{|c|c|c|c|}
 \multicolumn{4}{l}{\hspace{12mm}Types} \\ \hline
   1 & 2 & 3 & 4 \\ \hline
 $\star$ & $\geq$ 2 & $\star$ & $\star$ \\ \cline{1-4} 
  $\star$ & $\star$ & $\star$ &$\geq$ 2 \\ \cline{1-4} 
  $\star$ &  1 & $\star$ &  1 \\ \cline{1-4} 
  $\geq$ 1 &  1 &$\geq$ 1 & 0 \\ \cline{1-4} 
  $\geq$ 1 & 0 & $\geq$ 1 &  1 \\ \cline{1-4} 
  $\geq$ 2 & 0 & $\geq$ 2 & 0 \\ \hline
\end{tabular}

\label{Tab4_TwoC_Combo}
\end{table}

\paragraph{Broadcast Packet (BP)} \label{BP_4}
The list of type numbers for which ambiguity about activity or inactivity needs to be resolved in each block is informed to all nodes by the BS through a BP, which consists of a bit stream, just after the first phase (see Fig.~\ref{EST_Window_2}). 
For a given  $T$, let $S_{NS}(T)$ be the set of all  possible subsets of types of nodes for which the BS may  not be sure of their activity or inactivity for a block after the first phase; the case where the activity (or inactivity) of every type of node is known to the BS unambiguously is included in the set $S_{NS}(T)$ as an empty set ($\emptyset$). From Tables~\ref{Tab4_OneC} and~\ref{Tab4_TwoC_Combo}, we can see that $S_{NS}(4) = \{ \{3, 4\}, \{1, 2\}, \{1, 2, 3, 4\}, \emptyset \}$. Suppose the bit strings $00$, $01$, $10$ and $11$ are used to represent the cases $\emptyset, \{3, 4\}, \{1, 2\},  \{1, 2, 3, 4\}$ respectively.  The BS concatenates the bit strings corresponding to all the blocks of the first phase and sends the concatenated bit string in the BP; this concatenated list is received by all active nodes and they act accordingly during the second phase. For example (i) if the two bits in the $(2j-1)^{th}$ and $(2j)^{th}$ positions of the concatenated bit string are $00$, then it implies that the activity or inactivity of  nodes of all the types is known unambiguously to the BS for the $j^{th}$ block; hence all these nodes will be in silent mode during the second phase, (ii) if the two bits in the $(2j-1)^{th}$ and $(2j)^{th}$ positions of the concatenated bit string are $10$, then it implies that ambiguity exists regarding the activity or inactivity of nodes of Types 1 and 2 and the activity or inactivity of nodes of Types 3 and 4 has been inferred unambiguously for the $j^{th}$ block. Hence, during the second phase, active nodes of Types 1 and  2 corresponding to the $j^{th}$ block will participate to resolve the ambiguity and nodes of Types 3 and 4 will be in silent mode. 

Let $b(T)$ be the number of bits used to represent each element of the set $S_{NS}(T)$. Then the length of the BP (in terms of number of slots) for a given $T$ is given by:
\begin{equation}
\label{EQ_len_BP}
len_{BP}(T) = \ceil{(b(T) \times t_T)/S_W}.
\end{equation}
For $T=4$, $b(4) = \ceil{\log_2 |S_{NS}(4)|} = 2$. Hence, $len_{BP}(4) = \ceil{(2 t_4)/S_W}$. 

\paragraph{Second Phase} \label{SecondPhase_4}
In this phase only active nodes of those types and hash values (blocks) participate, for which ambiguity exists after the first phase. 
The first phase block results $C \beta$ and $\alpha C$ (see \Tabref{Tab4_OneC}) require one additional slot in the second phase to resolve the ambiguity. In case the result of a block is $C \beta$   (respectively, $\alpha C$), active Type 3 (respectively, Type 1) nodes corresponding to that block  respond with symbol $\alpha$ and active Type 4 (respectively, Type 2) nodes corresponding to that block respond with symbol $\beta$ in their corresponding slot of the second phase.  If the slot outcome is $\alpha$, then it implies that a node of Type 3 (respectively, Type 1) is active, else the slot outcome is $\beta$, and  it implies that a node of Type 4 (respectively, Type 2) is active. Thus, the ambiguity for the block is resolved at the end of the second phase. 
In case the  result  $C C$ occurs in a block of the first phase, then ambiguity exists regarding the activity or inactivity of nodes of all four types at the end of the first phase; then, for each of the groups (Type 1, Type 2) and (Type 3, Type 4), the estimation scheme for $T=2$ is used in the second phase to resolve this ambiguity. Note that Method II  is \emph{recursive}-- if the result $C C$ occurs in a block of the first phase while executing the scheme for $T = 4$, the scheme for $T = 2$ is used to resolve the ambiguity for that block in the second phase.



%

\subsubsection{Scheme for $T = 5$}\label{Method2_5}
The first phase consists of $t_5$ blocks, each consisting of two slots. 
From Fig.~\ref{Sym_Combo}, it follows that the symbol combinations given in the following table are used. 
\begin{center}
  \begin{tabular}{| c | c |}
    \hline
    Type & Symbol Combination \\ \hline
    1 & $\alpha$ $0$ \\ \hline
    2 & $\alpha$ $\alpha$ \\ \hline
    3 & $0$ $\beta$ \\ \hline
    4 & $\beta$ $\beta$ \\ \hline    
    5 & $\beta$ $\alpha$ \\ \hline
  \end{tabular}
\end{center}

\paragraph{First Phase} \label{FirstPhase_5}
The number of cases in which a collision occurs in exactly one slot of a  block  is $\binom{2}{1}\times3 = 6$.  The possible outcomes in a block in these cases and the inferences drawn by the BS regarding the activity or inactivity of various types of nodes  for each of these outcomes are shown in  \Tabref{Tab5_OneC}. 
From \Tabref{Tab5_OneC} it can be seen that the block results $C \alpha$, $C \beta$, $\alpha C$  and $\beta C$ require additional slots in the second phase to resolve ambiguity.  Finally, it is easy to check that if the result $C C$ occurs in a block, then ambiguity remains about the activity or inactivity of all five node types, and this ambiguity needs to be resolved in the second phase.  

\begin {table} [h]
\centering
\caption{{ This table shows the inferences drawn by the BS regarding the activity or inactivity of various types of nodes  for each of the outcomes in which  exactly one collision occurs per block.}}
\begin{tabular}{|c|c|c|c|c|} 
\hline
\multicolumn{2}{|c|}{Outcome in Block $B_i$} & \multicolumn{3}{c|}{Types} \\ \hline
Slot 1               & Slot 2              & Active  & Inactive    & Not Sure         \\ \hline
C                    & 0                   & 1     & 2, 3, 4, 5   & -                \\ \hline
C                    & $\alpha$            & 1   & 3, 4  & One of \{2, 5\}                 \\ \hline
C                    & $\beta$             & 1    & 2, 5   & One of \{3, 4\}   \\ \hline
0                    & C                   & 3    & 1, 2, 4, 5   & -                \\ \hline
$\alpha$             & C                   & 3   &  4, 5    & One of \{1, 2\}     \\ \hline
$\beta$              & C                   & 3   & 1, 2  & One of \{4, 5\}               \\ \hline
\end{tabular}

\label{Tab5_OneC}
\end{table}

\paragraph{Broadcast Packet (BP)} \label{BP_5}
A BP similar to that described in \Sectref{BP_4} is sent by the BS after the first phase. Using notation similar to that in that section,  $S_{NS}(5) = \{ \{2, 5\}, \{3, 4\}, \{1, 2\}, \{4, 5\}, \{1, 2, 3, 4, 5\}, \emptyset \}$. So $b(5) = \ceil{\log_2 |S_{NS}(5)|} = 3$ and $len_{BP}(5) = \ceil{b(5) \times t_5/S_W} = \ceil{3t_5/S_W}$.

\paragraph{Second Phase} \label{SecondPhase_5}
In this phase only active nodes of those types and hash values participate, for which ambiguity still exists after the first phase.  To resolve the  ambiguity in case of the block results $C \alpha$, $C \beta$,  $\alpha C$, and $\beta C$, an approach similar to the one described in \Sectref{SecondPhase_4} for the cases $C \beta$ and $\alpha C$ is used; in particular, one additional slot is required in the second phase. In case the block result  $C C$ occurs, the set of node types is divided into two groups: $\{1, 2, 3\}$ and $\{4, 5\}$. For the first (respectively, second) group, the scheme for $T = 3$ (respectively, $T=2$) is (recursively) used in the second phase to resolve the ambiguity.
%
%

\subsubsection{Scheme for $T = 6$}\label{Method2_6}
The first phase consists of $t_6$ blocks, each consisting of 3 slots. From Fig.~\ref{Sym_Combo}, it follows that the symbol combinations given in the following table are used.
\begin{center}
  \begin{tabular}{| c | c |}
    \hline
    Type & Symbol Combination \\ \hline
    1 & $\alpha$ $0$ $0$ \\ \hline
    2 & $\alpha$ $\alpha$ $0$ \\ \hline
    3 & $\alpha$ $\alpha$ $\alpha$ \\ \hline
    
    4 & $0$ $0$ $\beta$ \\ \hline
    5 & $0$ $\beta$ $\beta$ \\ \hline
    6 & $\beta$ $\beta$ $\beta$ \\ \hline
     \end{tabular}
\end{center}

\paragraph{First Phase} \label{FirstPhase_6}
The number of cases in which a collision occurs in exactly one slot (respectively, two slots) of a  block  is  $\binom{3}{1}\times3^2 = 27$ (respectively, $\binom{3}{2}\times3 = 9$). Out of these $27 + 9 = 36$ cases,  in \Tabref{Tab6_AllC}, we only show the cases where ambiguity about the activity or inactivity of some of the types of nodes exists and additional slots in the second phase are needed to resolve the ambiguity. Also, it is easy to check that if the result $C C C$ occurs in a block of the first phase, then ambiguity remains about the activity or inactivity of all six node types, and this ambiguity needs to be resolved in the second phase.

\begin {table}[h!]
\centering
\caption{{ This table shows the inferences drawn by the BS regarding the activity or inactivity of various types of nodes  for each of the outcomes in which  one or two collisions occur per block and ambiguity about the activity or inactivity of some of the types of nodes exists.}}
\begin{tabular}{|c|c|c|c|c|c|c|} 
\hline
\multicolumn{3}{|c|}{Outcome in Block $B_i$} & \multicolumn{3}{c|}{Types} \\ \hline
Slot 1      & Slot 2        & Slot 3       & Active & Inactive    & Not Sure         \\ \hline
C           & $\beta$       & $\beta$      & 1   & 2, 3, 4    & One of \{5, 6\}   \\ \hline
$\alpha$      & $\alpha$      & C          & 4    & 1, 5, 6   & One of \{2, 3\}   \\ \hline
C            & C            & 0            & 2    & 3, 4, 5, 6 & 1                   \\ \hline
C            & C            & $\alpha$     & 2, 3  & 4, 5, 6 & 1                   \\ \hline
C            & C            & $\beta$      & 2  & 3  & 1, One of \{4, 5, 6\} \\ \hline
0            & C            & C            & 5  & 1, 2, 3, 6  & 4                   \\ \hline
$\alpha$     & C            & C            & 5  & 6   & 4, One of \{1, 2, 3\} \\ \hline
$\beta$      & C            & C            & 5, 6  & 1, 2, 3 & 4                   \\ \hline
C            & $\alpha$     & C            & 1, 4  & 5, 6 & One of \{2, 3\}      \\ \hline
C            & $\beta$      & C            & 1, 4 & 2, 3 & One of \{5, 6\}      \\ \hline

\end{tabular}

\label{Tab6_AllC}
\end{table}

\paragraph{Broadcast Packet (BP)} \label{BP_6} 
A BP similar to that described in \Sectref{BP_4} is sent by the BS after the first phase. Using notation similar to that in that section, $S_{NS}(6) = \{ \{5, 6\},\{2, 3\}, \{1\}, \{1, 4, 5, 6\}, \{4\}, \{1, 2, 3, 4\}, \{1, 2, 3, 4, 5, 6\}$, $\emptyset \}$. So $b(6) = \ceil{\log_2 |S_{NS}(6)|} = 3$ and $len_{BP}(6) = \ceil{b(6) \times t_6/S_W} = \ceil{3t_6/S_W}$. 

\paragraph{Second Phase} \label{SecondPhase_6}
In this phase only nodes of those types and hash values participate, for which ambiguity still exists after the first phase. To resolve the ambiguity in case of the block results listed in  \Tabref{Tab6_AllC}, an  approach similar to those described in Sections~\ref{SecondPhase_4} and~\ref{SecondPhase_5} is used. For the block results  $C C \beta$ and $\alpha C C$, two slots are required in the second phase to resolve the ambiguity, whereas for the other block results, one slot is sufficient. For example, in case the block result $CC\beta$ occurs, in the first of the two required slots, each active Type 1 node transmits symbol $\alpha$. If the slot result is empty, it implies that all Type 1 nodes are inactive, else at least one Type 1 node is  active. In the second of the two required slots, each active Type 4 (respectively, Type 5, Type 6) node transmits symbol $\alpha$ (respectively, $\beta$, $0$ (no transmission)). Only three outcomes are possible: if the outcome is $\alpha$ (respectively, $\beta$, $0$), it implies that a Type 4 (respectively, Type 5, Type 6) node is active and all nodes of the other two types are inactive.  
Finally, in case the result  $C C C$ occurs in a block of the first phase, then the set of all node types is divided into two groups of size 3: $\{1, 2, 3\}$ and $\{4, 5, 6\}$, and the scheme for $T = 3$ is (recursively) used twice in the second phase to resolve the ambiguity.}

\section{Cognitive MAC Protocol for Multi-channel M2M Networks}\label{MAC_Proto}
In this section, for simplicity, we describe our proposed Cognitive MAC protocol for the case $T = 3$; the protocol in the case where $T$ is arbitrary is similar. Also, for concreteness, we assume that Type 1, Type 2 and Type 3 nodes are emergency, periodic and normal data type nodes respectively.
\subsection{Overview}
\begin{figure}[tbp]
  \centering
    \includegraphics[width=0.48\textwidth]{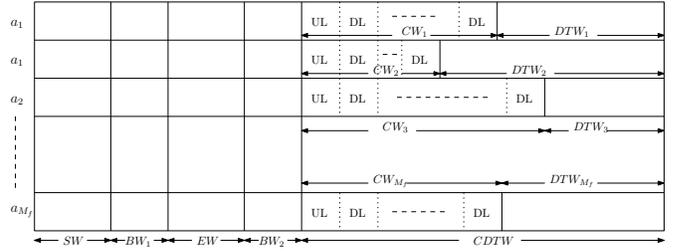}
    \caption{The figure shows the structure of a frame. Only the free channels are shown.}
    \label{Frame_Structure} 
\end{figure}

 Time is divided into frames of equal durations. Let $M_T$ be the total number of channels and $z_i$ be the probability that a primary user (PU) is present on channel $i \in \{1, \dots, M_T\}$ in any given frame~\footnote{We assume that the probabilities $z_i$ are known to the base station; for example, they can be estimated using past observations of PU occupancies on different channels.}. Also, in a frame, suppose there are $M_f$ free channels, say $\{a_1, a_2, \dots, a_{M_f}\}$; then out of these, we use the $M_{f_e}$ channels with the lowest values of $z_i$ for emergency data traffic, the $M_{f_p}$ channels with the next lowest values of $z_i$ for periodic data traffic and the rest for normal data traffic, for some $M_{f_e}, M_{f_p}$. We assume that all M2M devices are in the range of the base station (BS) (see Fig.~\ref{Network_Model}). In each frame, only the BS senses the $M_T$ channels to check for the presence of PUs. Note that since M2M devices are low-cost and battery-operated devices, our protocol does not require them to have sensing capability. 
 \Figref{Frame_Structure} shows the structure of a frame. The BS senses every channel in the sensing window (SW) to check for the presence of PUs. 
In the first broadcast window ($BW_1$), all the active nodes receive the list of channels that are free in the current frame from the BS (see \Sectref{BW1}). The fast node cardinality estimation scheme described in Section~\ref{EstSchemeMethod1} is executed in the estimation window (EW) to estimate the number of active nodes of each type~\footnote{Recall that for $T = 2$ and $T = 3$, the same estimation scheme is used under Methods I and II described in Section~\ref{EstSchemeforT}. For $T \geq 4$, either Method I or Method II can be used in the EW.} (see \Sectref{EW}). In the second broadcast window ($BW_2$), the list of channels assigned to each type of node and the optimal contention probabilities (which are computed using the estimates obtained in the EW) are broadcast by the BS (see \Sectref{BW2}). In the Contention and Data Transmission Window (CDTW), active nodes contend on the channels assigned to them using Slotted ALOHA~\cite{bertsekas1992data}; upon each successful contention, the BS reserves the requested number of slots for data transmission by the node in the DTW (see \Sectref{CDTW}). Periodic nodes require channels for periodically transmitting data. In particular, when a periodic node $r$ with $T_r$ data packets contends successfully, the BS reserves one slot each in $T_r$ successive frames for data transmissions by node $r$. Node $r$ does not participate again in the contention process in these $T_r$ frames. 

\subsection{First Broadcast Window ($BW_1$)}\label{BW1}
The BS and every node store the list of all channels, sorted in increasing order of $z_i$. In $BW_1$, the BS repeatedly broadcasts a packet on the first free channel (say $m_f$) of the above list; this packet contains the list of channels that are free in the current frame. 
Each active node tunes to channels in increasing order of $z_i$, listening for one time slot on each channel, until it tunes to channel $m_f$ and receives the list broadcast by the BS.
\subsection{Estimation Window (EW)}\label{EW}
\begin{figure}[tbp]
  \centering
\centering
\begin{tabular}{c|c|c|c|c|c|c|c|c} 
\cline{2-8}
$a_1$ & $1$ & $M_f + 1$ & . & . & . & . & . &  \\
\cline{2-8}
$a_2$ & $2$ & $M_f + 2$ & . & . & . & . & . &  \\
\cline{2-8}
$.$ & $.$ & . & . & . & . & . & . \\
$.$ & $.$ & . & . & . & . & . & $R_s - 1$ \\
$.$ & $.$ & . & . & . & . & . & $R_s$ \\
$.$ & $.$ & . & . & . & . & . & \\
$.$ & $.$ & . & . & . & . & . & \\
\cline{2-8}
$a_{M_f}$ & $M_f$ & $2M_f$ & . & . & . & . &  \\
\cline{2-8}
\end{tabular}
    \caption{The figure shows the scheme used for numbering the reserved $R_s$ slots in the EW. The first slot of channel $a_1$ is numbered $1$, 
the first slot of channel $a_2$ is numbered $2$, \ldots, the first slot of channel $a_{M_f}$ is numbered $M_f$, the second slot of channel $a_1$ is numbered $M_f + 1$ and so on.}
    \label{EW_SlotNumb} 
\end{figure}
Recall that the fast node cardinality estimation scheme described in Section~\ref{EstSchemeMethod1} requires $2t_3 + \ceil{t_3/S_W} + |C_I| + \ceil{|C_I|/S_W} + 2 |C_{II}|  = R_s$ (say) slots to execute. $R_s$ slots are reserved~\footnote{Note that although the value of $R_s$ is not known in advance, after the first (respectively, second) phase of the estimation scheme, the BS can find the value of $|C_I|$ (respectively, $|C_{II}|$) (see Sections~\ref{FirstPhaseforT} and~\ref{SecondPhaseforT}). So the information required to reserve $R_s$ slots is available with the network.} in the EW for the estimation process.  In the EW, \emph{all the $M_f$ free channels in the frame are utilized} for the estimation. The scheme used for numbering the reserved $R_s$ slots in the multi-channel environment is shown in \Figref{EW_SlotNumb}. 

\subsection{Second Broadcast Window ($BW_2$)}\label{BW2}
After the $EW$, the BS knows the estimated numbers of active nodes with emergency ($\hat{n}_e$), periodic ($\hat{n}_p$) and normal ($\hat{n}_n$) data  packets. Based on the values of $\hat{n}_e$, $\hat{n}_p$ and $\hat{n}_n$, out of the $M_f$ free channels, $M_{f_e}$, $M_{f_p}$ and $M_{f_n}$ channels are assigned to emergency, periodic and normal data nodes respectively, where  $M_f = M_{f_e} + M_{f_p} + M_{f_n}$; the BS broadcasts the lists of channels assigned to each type of node in $BW_2$. For instance, let $w_{e}, w_{p}$ and $w_{n}$ be  \emph{weights} (positive real numbers) associated with the emergency,   periodic and normal data classes respectively.  
Then $M_{f_e}, M_{f_p}$ and  $M_{f_n}$ may be selected to be approximately  $\frac{\hat{n}_e w_{e} M_f}{(\hat{n}_e w_{e} + \hat{n}_p w_{p} + \hat{n}_n w_{n})}$, $\frac{\hat{n}_p w_{p} M_f}{(\hat{n}_e w_{e} + \hat{n}_p w_{p} + \hat{n}_n w_{n})}$ and $\frac{\hat{n}_n w_{n} M_f}{(\hat{n}_e w_{e} + \hat{n}_p w_{p} + \hat{n}_n w_{n})}$ respectively. 
We use $w_{e} \geq w_{p} \geq w_{n}$ to ensure that emergency (respectively, periodic) data is provided a higher priority than periodic (respectively, normal) data. To balance the load across the assigned channels, each emergency  (respectively, periodic, normal) node selects one channel from the $M_{f_e}$ (respectively, $M_{f_p}$, $M_{f_n}$) channels uniformly at random and tunes to it in the CDTW. Now, recall that if $n$ nodes contend using Slotted ALOHA, then the value of the contention probability $p$ that maximizes the throughput is $p = {1}/{n}$~\cite{bertsekas1992data}. So the BS sets the probabilities of contention for emergency, periodic and normal data nodes  to $\hat{p}_e = \min({M_{f_e}}/{\hat{n}_{e}}, 1)$, $\hat{p}_p = \min({M_{f_p}}/{\hat{n}_{p}}, 1)$ and $\hat{p}_n = \min({M_{f_n}}/{\hat{n}_{n}}, 1)$ respectively and broadcasts the values of $\hat{p}_e$, $\hat{p}_p$ and $\hat{p}_n$ in $BW_2$. Finally, there may be some periodic data nodes with time slots in the DTW of the current frame reserved in past frames; a packet containing a list of such reserved slots is also broadcast by the BS in $BW_2$. 
 
\subsection{Contention and Data Transmission Window (CDTW)}\label{CDTW}
After $BW_2$, all nodes switch to their respective selected channels for contention and data transmission. Every channel in this window is divided into a Contention Window (CW) and a Data Transmission Window (DTW) of variable lengths (see \Figref{Frame_Structure}). Each active node from $\mathcal{N}_1$ contends using Slotted ALOHA~\cite{bertsekas1992data} with contention probability $\hat{p}_e$ in the first slot of the CW on its channel, which is an uplink ($UL$) slot, and listens to the channel for an acknowledgment (ACK) packet from the BS in the second slot, which is a downlink ($DL$) slot.  If a node successfully contends in the $UL$ slot, then it is allotted the requested number of slots in the DTW by the BS, starting from the rightmost available slot of the current frame; the BS includes the list of allotted slots in the ACK packet that it broadcasts in the following $DL$ slot. The node then switches to idle (sleep) state to save energy and wakes up to transmit in its allotted slots in the DTW. Now, since the number of contending nodes has reduced by $1$, the BS modifies $\hat{p}_e$ to $\min(1/({(\hat{n}_{e}}/{M_{f_e}}) - 1), 1)$ and broadcasts this value in the $DL$ slot. In case of an unsuccessful contention (collision or empty slot), the BS does not send any ACK. This process continues until the CW and DTW on that channel are separated by a single slot; then, the BS transmits a broadcast packet informing the remaining contending nodes to switch to idle state (to save energy) for the rest of the frame. However, if three successive $UL$ slots are empty, then it is taken by the BS to be an indication that with a high probability all the active nodes on the channel have already successfully contended; in this case, the BS can allot the remaining free slots of the channel to active nodes of other channels.  A similar procedure is followed by active nodes from $\mathcal{N}_2$ and $\mathcal{N}_3$ on their selected channels with parameter sets $(\hat{p}_p, \hat{n}_p)$ and $(\hat{p}_n, \hat{n}_n)$ respectively.  

\section{Performance Analysis}\label{Ana_Res}
In this section, we obtain closed-form expressions for  the expected number of successful contentions per frame and the expected amount of energy consumed in the CDTW of a frame  under the Cognitive  MAC protocol described in Section~\ref{MAC_Proto}. 
\begin{figure}[tbp]
  \centering
    \includegraphics[width=0.48\textwidth]{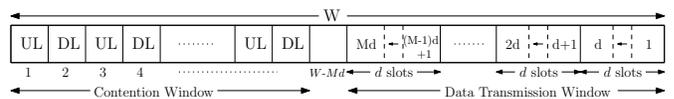}
    \caption{The figure shows the CDTW used in the analysis in Section~\ref{SSC:expected:M}.}
    \label{Ana_Model} 
\end{figure}

\subsection{Expected Number of Successful Contentions}
\label{SSC:expected:M}
Here, we focus on only one channel and hence only nodes of a single type contend on it. Assume that $n$ nodes of this type are active on the channel at the start of a given frame and let $\hat{n}$ be the estimated value of $n$. Let the length of the CDTW of the frame be $W$ slots.

Let $M$ be the number of successful contentions in the given frame. Recall that contentions occur only in $UL$ slots. For tractability, we assume that upon every successful contention, the BS reserves a constant number, say $d$, of slots for the successful node from the last available slot in that frame as shown in \Figref{Ana_Model}. If no successful contentions take place in the frame, then $M=0$ and if all contentions are successful, then  $M= \lfloor {W}/{(2+d)} \rfloor$ since $2 M + dM = W$ or $W-1$. So $0 \le M \le {W}/{(2+d)}$. For each $x$, let $W_x = \lfloor\frac{1}{2} (W - xd)\rfloor$. By definition: 
\begin{equation}
\label{Exp_M}
E(M) = \sum_{m=0}^{\lfloor W/(2+d) \rfloor} m P(M=m) 
\end{equation}
Let $S_x$ be the event that a successful contention occurs when $x$ nodes contend. Let $P(S_x) = r_x$ and $p_x$ be the contention probability used when $x$ nodes contend. In the proposed protocol, $p_{n-j} = \min({1}/({\hat{n} - j}), 1)$, $j = 0, 1, 2 \ldots$ (see Section~\ref{CDTW}). We now find the distribution of $M$. There are $M=m$ successful contentions if and only if for some integers $k_1, k_2, \ldots, k_m$, the first $(k_1 - 1)$ contention attempts are unsuccessful with $n$ contending nodes and the $k_1^{th}$ attempt is successful, $(k_1 + 1)^{th}$ to $(k_2 - 1)^{th}$ attempts are unsuccessful with $n- 1$ contending nodes and $k_2^{th}$ attempt is successful, \ldots, and $(k_{m} + 1)^{th}$ to $W_m ^{th}$ attempts are unsuccessful with $n- m$ contending nodes. So: 
\begin{eqnarray}
\label{PMm}
P(M=m) =  \sum_{k_1=1}^{W_m - m + 1} \sum_{k_2=k_1 + 1}^{W_m - m + 2} \ldots \sum_{k_{j+1}=k_j + 1}^{W_m - m + j + 1}  \ldots \sum_{k_{m}=k_{m-1} + 1}^{W_m}  \nonumber \\(1 - r_n)^{k_1 - 1}r_n (1 - r_{n-1})^{k_2 - k_1 - 1}r_{n-1} \ldots \nonumber \\ (1 - r_{n-j})^{k_{j+1} - k_j - 1}r_{n-j} \ldots (1 - r_{n-m})^{W_m - k_m}.
\end{eqnarray}
Note that $r_{n-j} = (n- j) p_{n-j} (1 - p_{n-j})^{(n- j - 1)}$. $E(M)$ can be obtained from~\eqref{Exp_M} and~\eqref{PMm}.

\subsection{Expected Amount of Energy Consumed in the CDTW of a Frame}
Let $\gamma_I, \gamma_T, \gamma_R$ be the energy spent by a node per slot in the idle state, transmission state and reception state respectively. Let us classify the slots in the CDTW of a given frame into uplink slots, downlink slots and data transmission slots; let the total energy spent by all the active  nodes in them be  $\mathscr{E}_{UL}, \mathscr{E}_{DL}$ and $\mathscr{E}_{DT}$ respectively. So the total expected amount of energy spent in the CDTW of a frame is $E(\mathscr{E}_{UL}) + E(\mathscr{E}_{DL}) + E(\mathscr{E}_{DT})$. We compute $E(\mathscr{E}_{UL})$, $E(\mathscr{E}_{DL})$ and $E(\mathscr{E}_{DT})$ in Sections~\ref{EUL},~\ref{EDL} and~\ref{EDT} respectively.

\subsubsection{Energy Spent in UL Slots}\label{EUL}
Note that there are a total of $W_M$ uplink slots in the frame. In each of these slots, some of the active nodes are in transmission state and the rest are in idle state. So, $\mathscr{E}_{UL} = \sum_{i=1}^{W_M} \big(L_i \gamma_T + (n-L_i)\gamma_I\big)$, 
where $L_i$ is the number of nodes that transmit in $UL$ slot $i$, which depends on $N_i$ (the number of contending nodes in $UL$ slot $i$) and $p_{N_i}$. Taking expectations and conditioning on the values taken by $M$:
\begin{multline}
\label{Exp_EUL}
E(\mathscr{E}_{UL}) =  \sum_{m=0}^{ \lfloor W/(2+d) \rfloor} \Bigg( \sum_{i=1}^{W_m} \bigg(E(L_i/M=m)  \gamma_T \\ +  \Big(n- E(L_i/M=m)\Big)\gamma_I \bigg)\Bigg) P(M=m).
\end{multline}
{ Next, we find an expression for $E(L_i/M=m)$ in closed form; $E(\mathscr{E}_{UL})$ can be found using this expression,~\eqref{PMm} and~\eqref{Exp_EUL}. 
Now, $E(L_i/M=m)$ can be calculated as:
\begin{multline}
\label{ELi} 
E(L_i/M=m)  = \sum_{j=0}^m \Big(\sum_{l_i = 0}^{n-j} l_i P(L_i = l_i/M=m, N_i = n-j)\Big)\\ P(N_i = n-j/M=m)
\end{multline}

$P(N_i = n-j/M=m)$ for all $i$ and $j$ can be found using the fact that $N_1 = n$ with probability $1$ and the following recursion: 
\begin{multline}
\label{Cond_Pro}
P(N_i = n-j/M=m) = P(N_{i-1} = n-j/M=m) \\ \Big(1 - P_{i-1}(S_{n-j}/M=m)\Big)  + P(N_{i-1} = n-j+1/M=m)\\\Big(P_{i-1}(S_{n-j+1}/M=m)\Big)
\end{multline}
where $P_i(S_{x}/M=m)$ is the probability of success when $x$ nodes contend in UL slot $i$ given $M=m$ and $j \in \{0, 1, \dots ,i-1\}$. 
Now,
\begin{equation}
\label{PSNj} 
P_i(S_{n-j}/M=m) = \frac{P_i(M=m/S_{n-j}) P_i(S_{n-j})}{P(M=m)},
\end{equation}
where $P_i(S_{x}) = x p_x (1 - p_x)^{x-1}$ is the probability of success when $x$ nodes contend in UL slot $i$ and $P_i(M=m/S_{x})$ is the probability that $M = m$ given that a success occurs when $x$ nodes contend in UL slot $i$. Next, similar to \eqref{PMm}, we get: 
\begin{eqnarray}
P_i(M=m/S_{n-j}) = \sum_{k_1=i+1}^{W_m - m + 1} \sum_{k_2=k_1 + 1}^{W_m - m + 2}    \dots \sum_{k_{m}=k_{m-1} + 1}^{W_m}  \nonumber \\(1 - r_{n-j-1})^{k_1 - i - 1}r_{n-j-1} (1 - r_{n-j-2})^{k_2 - k_1 - 1}r_{n-j-2} \nonumber \\ \dots (1 - r_{n-m})^{W_m - k_m}.
\label{PiMm}
\end{eqnarray}
for $m \geq j+1$ and $P_i(M=m/S_{n-j}) = 0$ for $m < j+1$. Next:
\begin{dmath}
 P(L_i = l_i | M = m, N_i = n-j)
 =   \frac{P(L_i = l_i, M = m| N_i = n-j)}{P(M=m|N_i = n-j)} \nonumber \\
  =  \frac{P(M=m|L_i = l_i, N_i = n-j)P(L_i = l_i| N_i = n-j)}{P(M=m|N_i = n-j)}
\label{EQ:conditional:PLi}
\end{dmath}
Note that closed form expressions for $P(M=m|L_i = l_i, N_i = n-j)$ and $P(M=m|N_i = n-j)$ can be computed similar to the computation of $P_i(M=m/S_{n-j})$ in \eqref{PiMm}. Also, clearly:
\begin{equation}
P(L_i = l_i| N_i = n-j) = \left( \begin{array}{c}
n-j \\
l_i \\
\end{array}
\right) p_{n-j}^{l_i} (1 - p_{n-j})^{n-j-l_i}. 
\label{EQ:PLi:Ni:conditional}
\end{equation}

Finally, by using~\eqref{PMm}$-$\eqref{EQ:PLi:Ni:conditional}, 
we get $E(\mathscr{E}_{UL})$.}

\subsubsection{Energy Spent in DL Slots}\label{EDL}
In these slots, contending nodes are in reception state and the rest are in idle state. So $\mathscr{E}_{DL} = \sum_{i=1}^{W_M} \big(N_i \gamma_R + (n-N_i)\gamma_I\big)$. 
Taking expectations and conditioning on the values taken by $M$, we get:
\begin{multline}
\label{Exp_EDL}
E(\mathscr{E}_{DL}) =  \sum_{m=0}^{\lfloor W/(2+d) \rfloor} \Bigg( \sum_{i=1}^{W_m} \bigg(E(N_i/M=m)  \gamma_R \\ + \Big(n-E(N_i/M=m)\Big)\gamma_I \bigg)\Bigg) P(M=m).
\end{multline}
By definition of conditional expectation, $E(N_i/M=m) = \sum_{j=0}^m (n-j) P(N_i = n-j/M=m)$. Hence, $E(\mathscr{E}_{DL})$ can  be found using~\eqref{PMm},~\eqref{Cond_Pro} and~\eqref{Exp_EDL}.

\subsubsection{Energy Spent in Data Transmission Slots}\label{EDT}
Since only one node is transmitting in these slots, all other nodes are in the idle state. So $\mathscr{E}_{DT} = d M \Big(\gamma_T + (n-1)\gamma_I\Big)$ and 
\begin{equation}
\label{Equ_DT}
E(\mathscr{E}_{DT}) = d E(M) \Big(\gamma_T + (n-1)\gamma_I\Big).
\end{equation}
Using \eqref{Exp_M} and \eqref{Equ_DT}, we can calculate $E(\mathscr{E}_{DT})$.

\section{Simulations}\label{Simu}
In this section, via simulations~\footnote{All our simulations were done using the MATLAB software.}, we evaluate: (i) the performances of the node cardinality estimation schemes proposed in Sections~\ref{EstSchemeMethod1} and~\ref{EstSchemeMethod2} (see Section~\ref{Simu1}), and (ii) the performance of the proposed MAC protocol described in \Sectref{MAC_Proto} (see Section~\ref{Simu2}).

{ \subsection{Performances of the Node Cardinality Estimation Schemes}\label{Simu1}}
{ In this subsection, we compare the performances of the proposed estimation schemes, viz., Method I (described in~\Sectref{EstSchemeMethod1}) and Method II (described in~\Sectref{EstSchemeMethod2}) with that of a protocol in which the LoF based protocol~\cite{qian2011cardinality} is executed $T$ times to separately estimate the active node cardinalities of the $T$ types of nodes. Throughout this subsection, we assume that $|\mathscr{N}_1| = |\mathscr{N}_2| = \ldots = |\mathscr{N}_T| = D$ (say) and $n_{1, all} = n_{2, all} = \ldots = n_{T, all} = n_{all}$ (say). Let $q_i$ (respectively,  $1 - q_i$) be the probability with which a given  node of Type $i$ is active (respectively,  inactive) in a given time frame. 

\begin{figure}[ht]
  \centering  
  \begin{minipage}[b]{0.5\linewidth}
        \centering
        \resizebox{1.0\columnwidth}{!}{\includegraphics{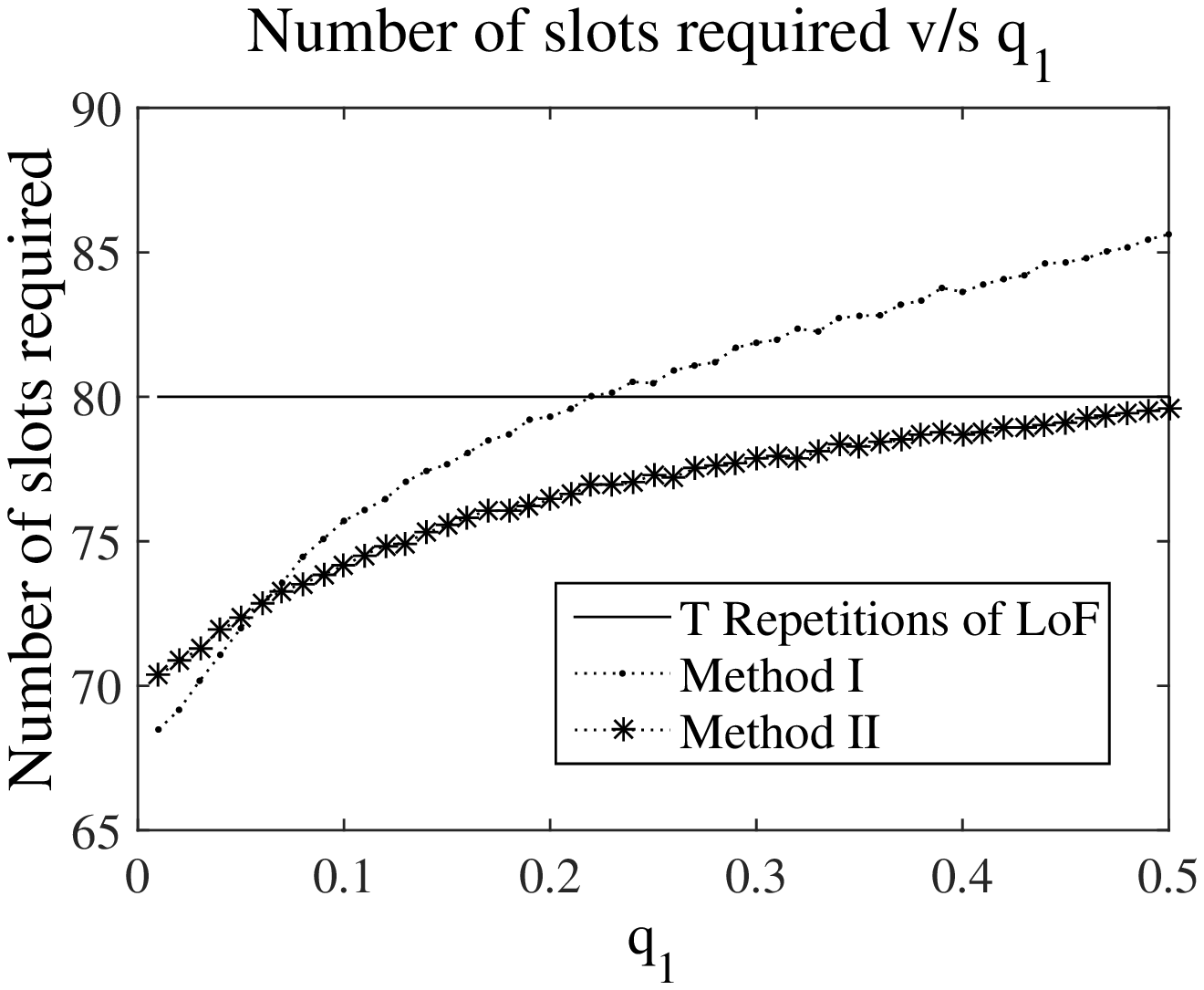}}
    	\label{Het_q1} 
\end{minipage}%
\begin{minipage}[b]{0.5\linewidth}
  \centering
  \resizebox{1.0\columnwidth}{!}{\includegraphics{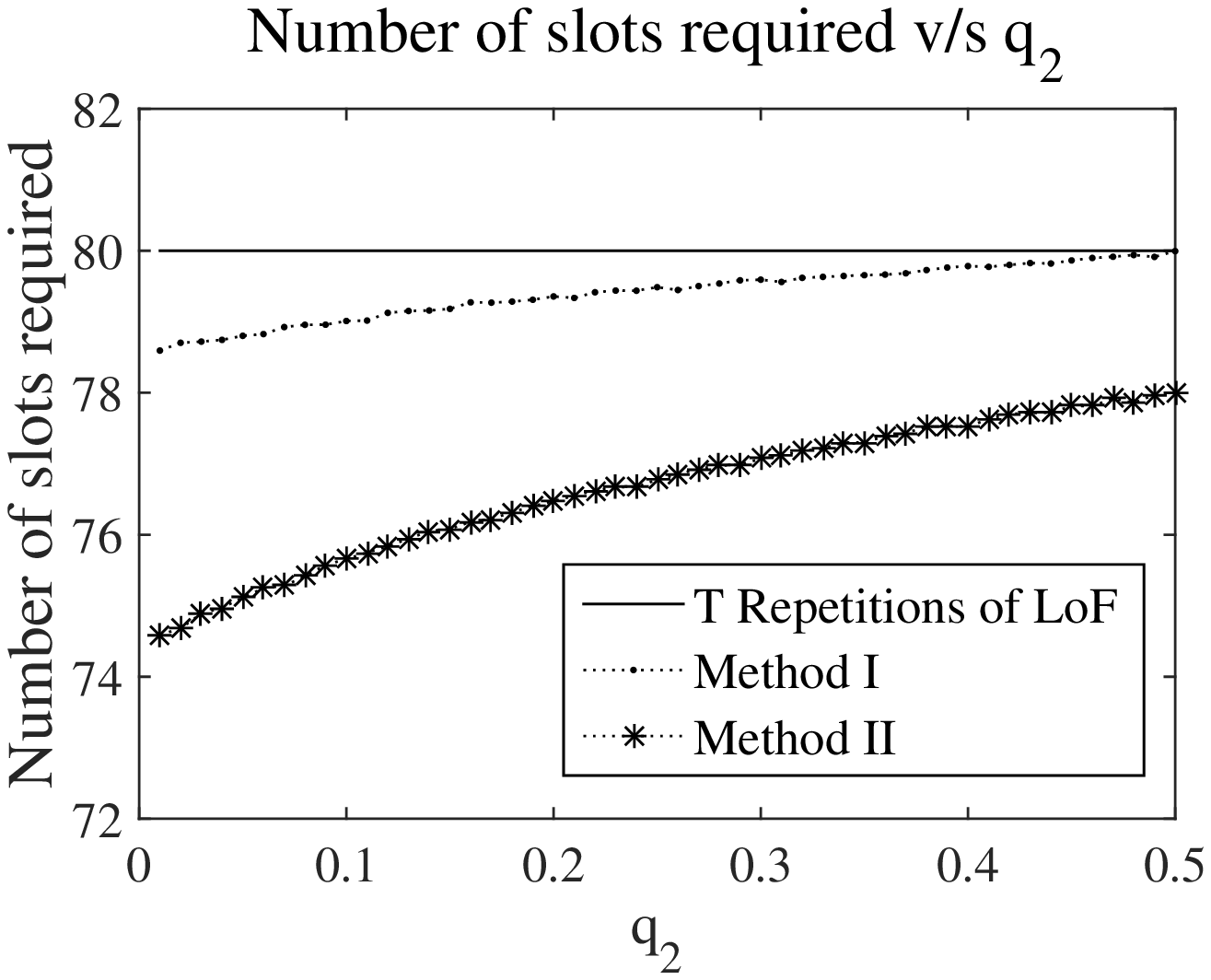}}
     \label{Het_q2} 
\end{minipage}
    \caption{{ The following parameters are used in both plots: $T = 4$, $D = 100$,  $q_3 = 0.75$, $q_4 = 0.12$. Also, $q_2 = 0.2$ (respectively, $q_1 = 0.2$) is used in the left (respectively, right) plot.}}
    \label{Het_Case}
\end{figure}

The  left (respectively, right) plot of \Figref{Het_Case} shows the number of slots required for Method I, Method II and the $T = 4$ repetitions of the LoF based protocol to execute versus $q_1$ (respectively,  $q_2$).  Both plots of \Figref{Het_Case} show that for sufficiently low values of the probability that a given node is active ($q_1$ or $q_2$), both the proposed methods, Method I and Method II, outperform the $T$ repetitions of LoF based protocol. Intuitively, this is because in case of Method I and Method II, as the probability that a given node is active increases, so does the number of collisions in the first phase and/ or second phase, and hence the total number of slots required to execute.  Thus, in M2M networks in which the nodes are, e.g., sensors that occasionally transmit measurements or nodes that transmit alarms, emergency alerts and other infrequent messages, the two proposed protocols are expected to perform well.

\begin{figure}[ht]
  \centering  
  \begin{minipage}[b]{0.5\linewidth}
        \centering
        \resizebox{1.0\columnwidth}{!}{\includegraphics{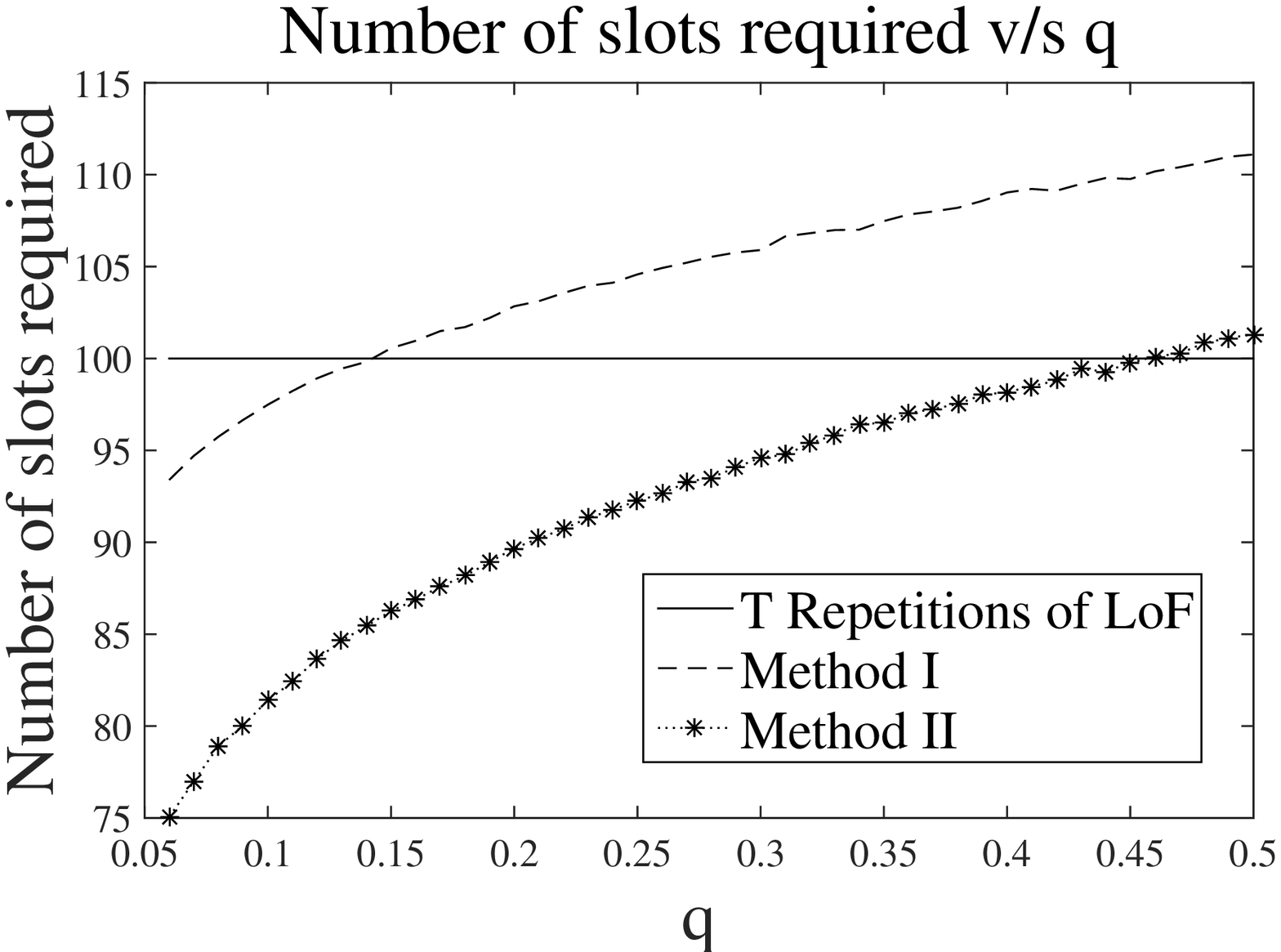}}
    	\label{q_T5} 
\end{minipage}%
\begin{minipage}[b]{0.5\linewidth}
  \centering
  \resizebox{1.0\columnwidth}{!}{\includegraphics{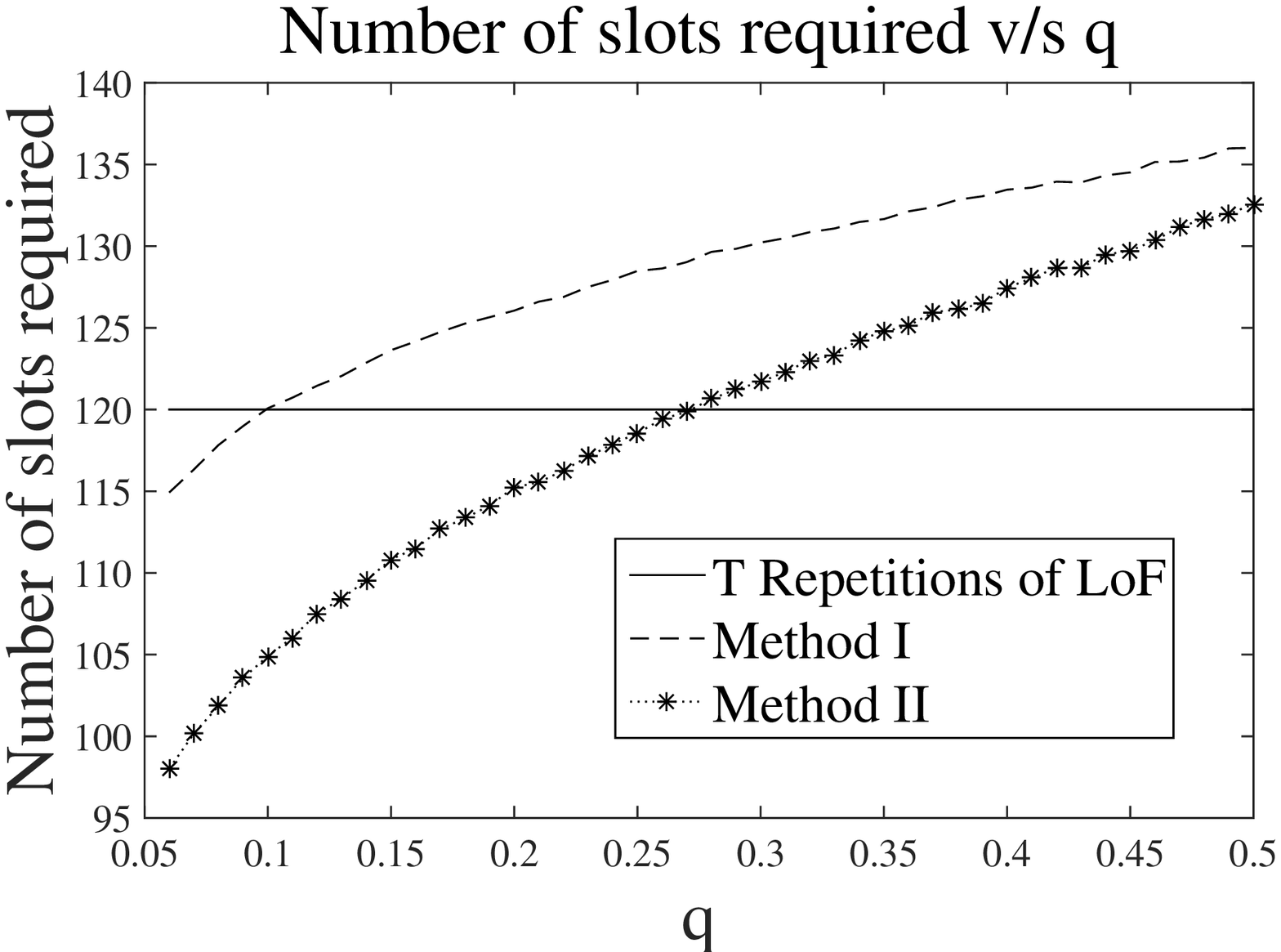}}
     \label{q_T6} 
\end{minipage}
    \caption{{ $D = 100$ is used in both plots. $T=5$ (respectively, $T=6$) is used in the left (respectively, right) plot. From these plots, we can deduce that $q^I_{threshold} = 0.13, q^{II}_{threshold} = 0.45$ for $T=5$ and $q^I_{threshold} = 0.10, q^{II}_{threshold} = 0.27$ for $T=6$.}}
    \label{Vs_q}
\end{figure}

The left (respectively, right) plot of \Figref{Vs_q} shows the number of slots required for $T=5$ (respectively,  $T=6$) by Method I, Method II and the $T$ repetitions of LoF based protocol to execute versus $q$ in the case where $q_i = q$,  $\forall i$. Again, both plots show that for sufficiently low values of $q$, both Method I and Method II outperform the $T$ repetitions of LoF based protocol. Also, these plots show that Method II outperforms Method I for all the values of $q$ considered; thus, the fact that Method II is more sophisticated than Method I leads to better performance of the former than the latter.  Let $q^I_{threshold}$ (respectively, $q^{II}_{threshold}$) be the value of $q$ below which Method I (respectively, Method II) performs better than the $T$ repetitions of LoF based protocol. 

\begin{figure}[ht]
  \centering  
  \begin{minipage}[b]{0.5\linewidth}
        \centering
        \resizebox{1.0\columnwidth}{!}{\includegraphics{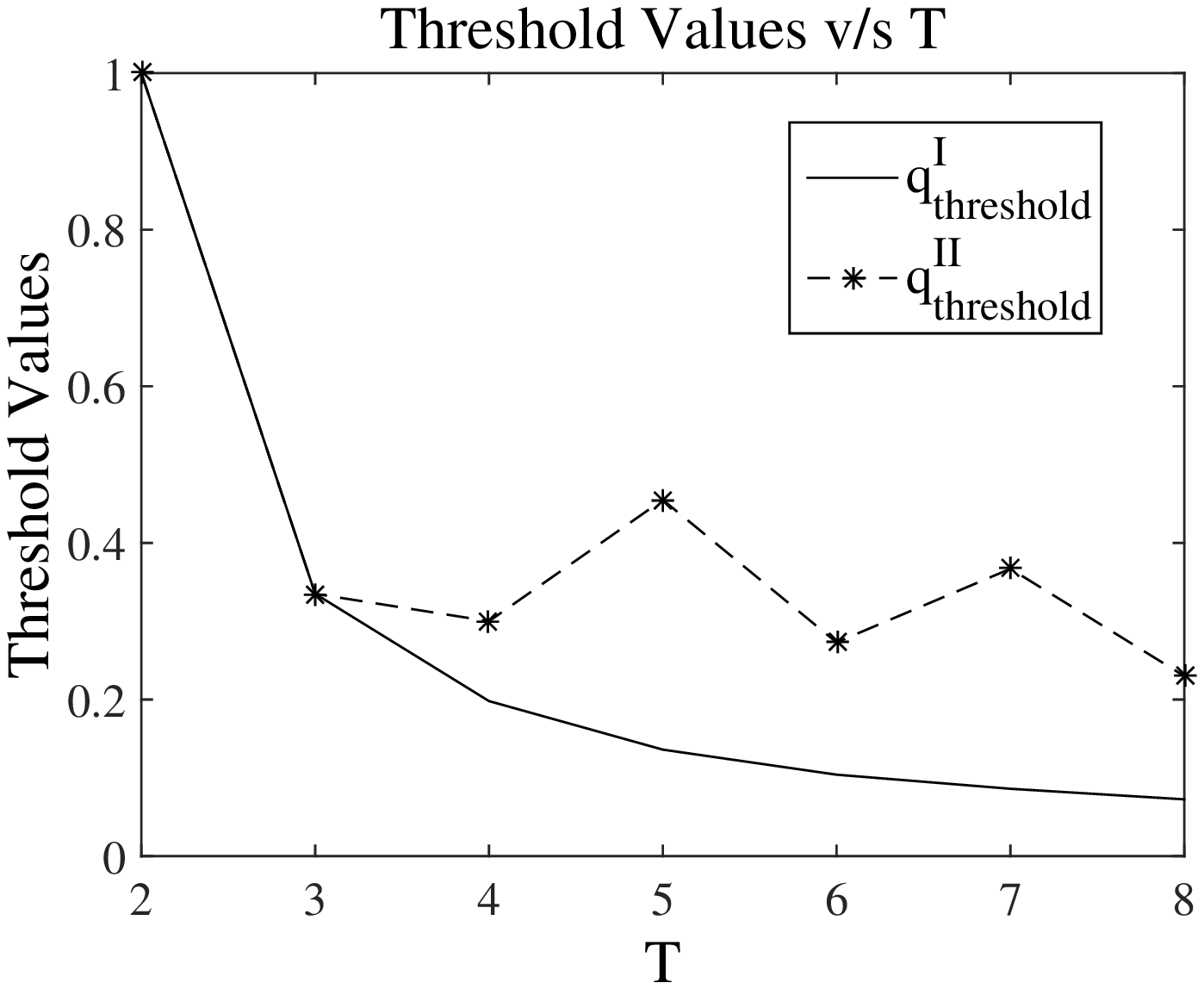}}
    	\label{q_stable_vs_T} 
\end{minipage}%
\begin{minipage}[b]{0.5\linewidth}
  \centering
  \resizebox{1.0\columnwidth}{!}{\includegraphics{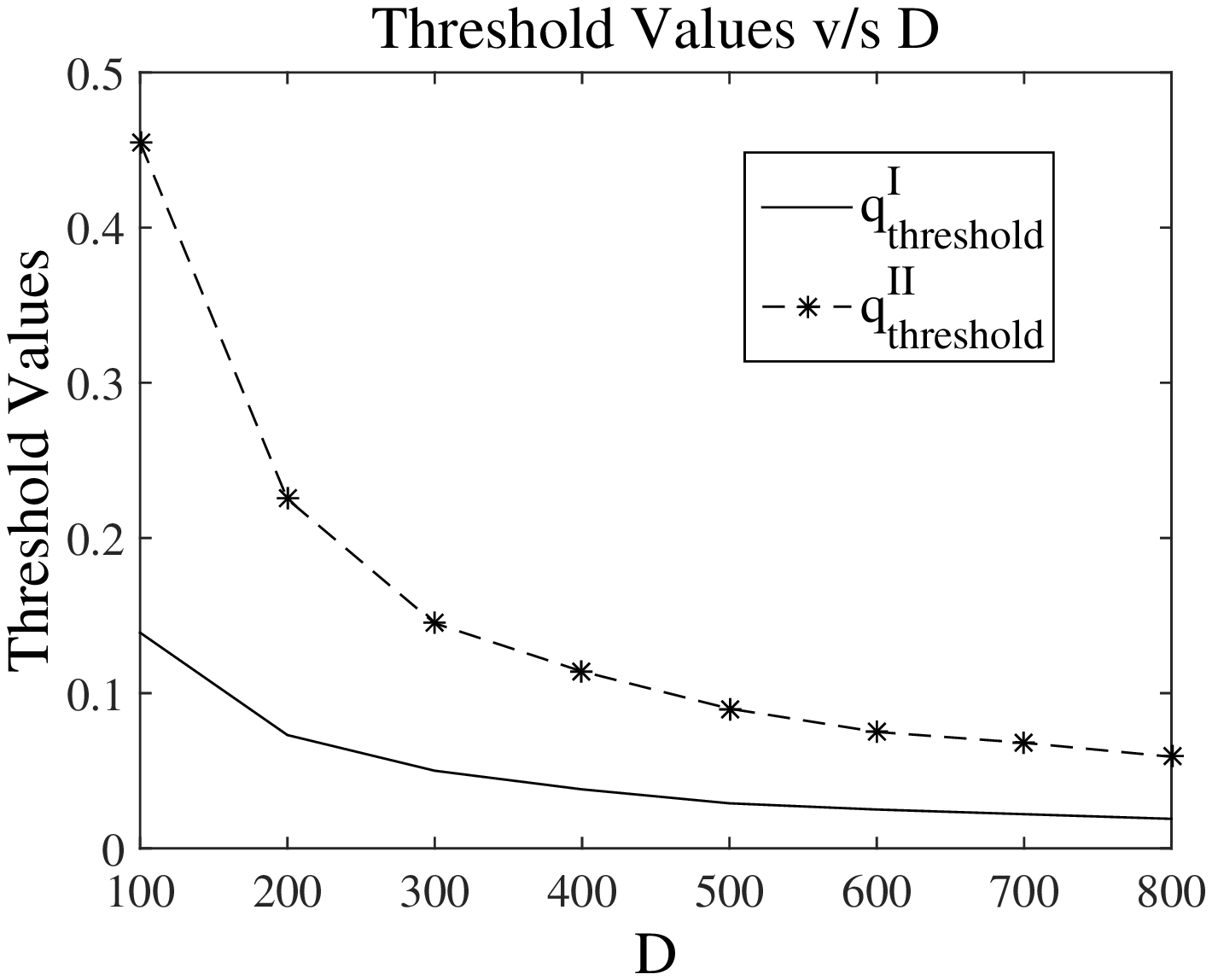}}
     \label{q_stable_vs_N} 
\end{minipage}
    \caption{{ $D = 100$ (respectively, $T = 4$) is used in the left (respectively, right) plot.}}
    \label{Vs_T_N}
\end{figure}

For the case where $q_i = q$, $\forall i$, the left (respectively, right) plot of \Figref{Vs_T_N} shows the threshold values $q^I_{threshold}$ and  $q^{II}_{threshold}$ versus $T$ (respectively, $D$). Again, both plots show that Method II performs better than Method I~\footnote{{ Note that for $T = 2$ and $T= 3$, we use the same scheme in both the methods. Hence, $q^I_{threshold}= q^{II}_{threshold}$ for $T = 2$ and $T= 3$. Also, the left plot of \Figref{Vs_T_N} shows that for odd values of $T \geq 5$, $q^{II}_{threshold}$ is higher than $q^{II}_{threshold}$ for $T-1$ (which is even).  Intuitively, this is because in Method II, for even values of $T$, $(T/2)t_T$ slots are used in the first phase, whereas for odd values of $T$, only $((T-1)/2)t_T$ slots are used in the first phase.}.} The right plot shows that both $q^I_{threshold}$ and $q^{II}_{threshold}$ decrease in $D$; this is because as the number of nodes in the network increases, more collisions occur in the first and/ or second phase of Methods I and II, which results in an increase in the number of slots they require to execute.

\begin{figure}[ht]
  \centering  
  \begin{minipage}[b]{0.5\linewidth}
        \centering
        \resizebox{1.0\columnwidth}{!}{\includegraphics{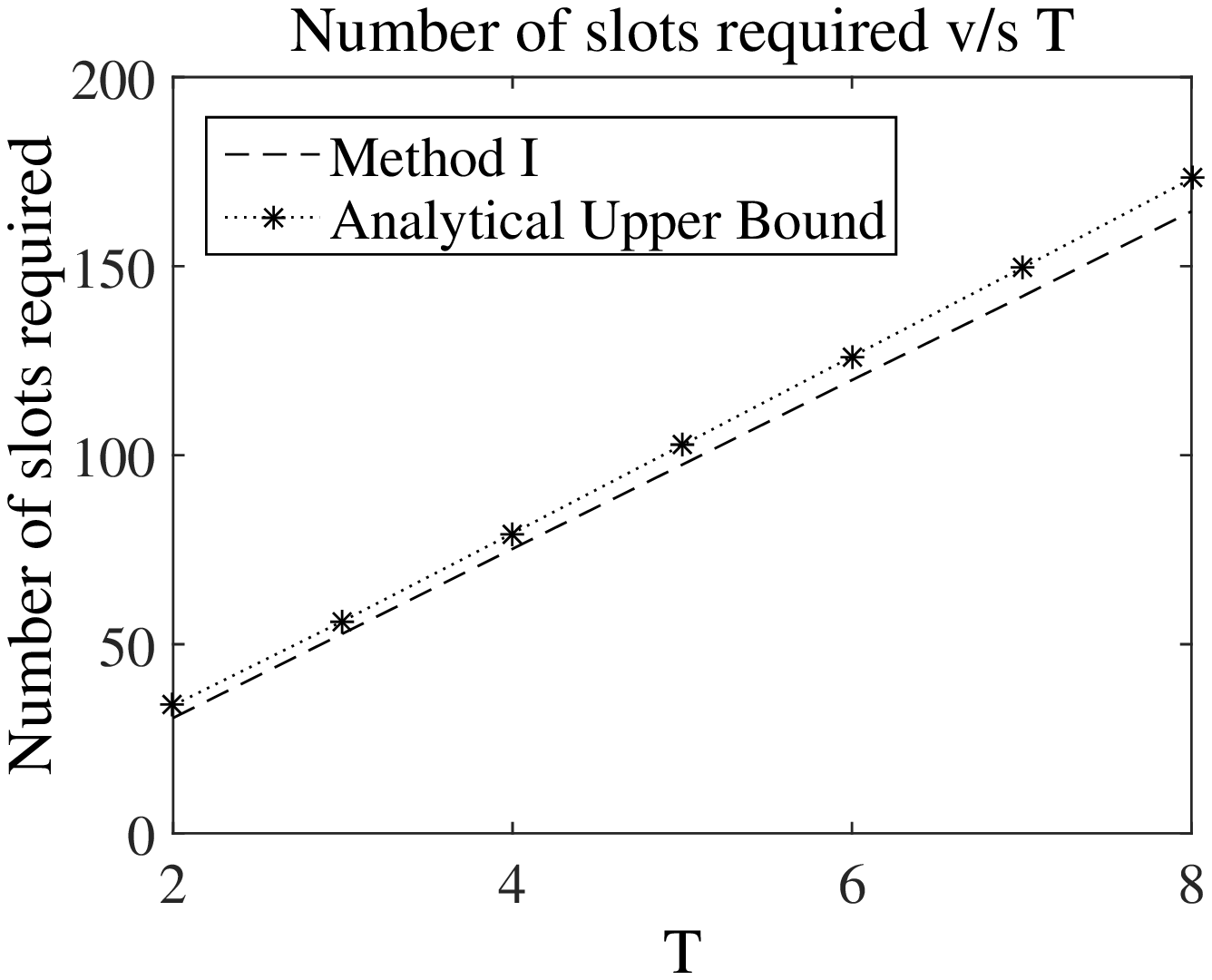}}
    	\label{UB_q03} 
\end{minipage}%
\begin{minipage}[b]{0.5\linewidth}
  \centering
  \resizebox{1.0\columnwidth}{!}{\includegraphics{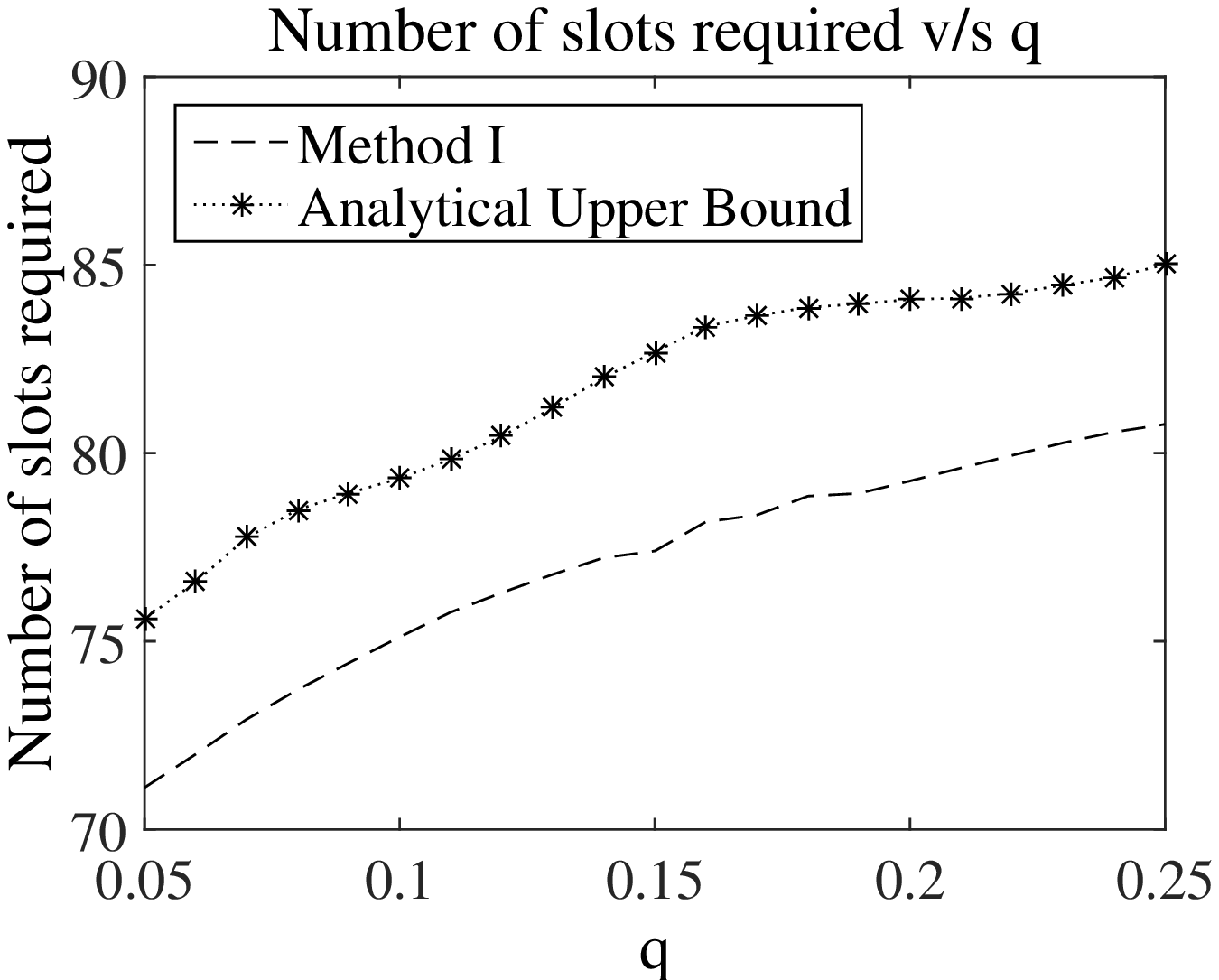}}
     \label{UB_T4} 
\end{minipage}
    \caption{{ $D = 100$ is used in both the plots. $q = 0.1$ (respectively, $T = 4$) is used in the left (respectively, right) plot.}}
    \label{Compare_UB}
\end{figure}

The left (respectively, right) plot of \Figref{Compare_UB} shows the analytical upper bound on the expected number of slots required (derived in \Sectref{UB_EstforT}) and the exact expected number of slots required by Method I to execute versus $T$ (respectively, $q$) in the case where $q_i = q$, $\forall i$. These plots show that the upper bound is fairly tight.}

\subsection{Performance of the Proposed Cognitive MAC Protocol}\label{Simu2}
In this subsection, we evaluate the performance of the proposed Cognitive MAC protocol described in \Sectref{MAC_Proto}, in terms of average throughput and average delay, via simulations. Also, we compare the performance of the proposed protocol with a hypothetical ``ideal protocol'' to find out how accurate the proposed estimation scheme is. The ideal protocol is similar to the proposed protocol, with the difference being that it is assumed to know the \emph{exact} number of active nodes at any time~\footnote{Note that the ideal protocol is not practically implementable and is considered only for comparison with the proposed protocol.}.

Let $M_T, M_f, \hat{n}_e, \hat{n}_p, \hat{n}_n, w_{e}, w_{p},  w_{n}$ and $z_i$ be as defined in \Sectref{MAC_Proto}. At the beginning of each frame, data packets arrive at random at each node; the number of packets that arrive at a node belonging to  the emergency (respectively, periodic, normal) data class is a Poisson random variable  with mean $\lambda_{e}$ (respectively,  $\lambda_{p}$,  $\lambda_{n}$). Also, each frame is divided into $50$ slots and transmission of a data packet takes $1$ slot.  An emergency (respectively, normal) node which has successfully contended for access during the CW can reserve at most $k_{e}$ (respectively, $k_{n}$) consecutive slots in the DTW. A periodic node can reserve one slot per frame for at most $k_{p}$ consecutive frames. The limits $k_{e}$, $k_{n}$ and $k_{p}$ are imposed to ensure short-term fairness in the transmission opportunities that different nodes get.
We consider a balanced load condition wherein there are an equal number, say $N$, of nodes of each type in the network and  $\lambda_{e}=\lambda_{p}=\lambda_{n}=\lambda$ (say). 

\begin{figure}[t!]
  \centering  
  \begin{minipage}[b]{0.5\linewidth}
        \centering
        \resizebox{1.0\columnwidth}{!}{\includegraphics{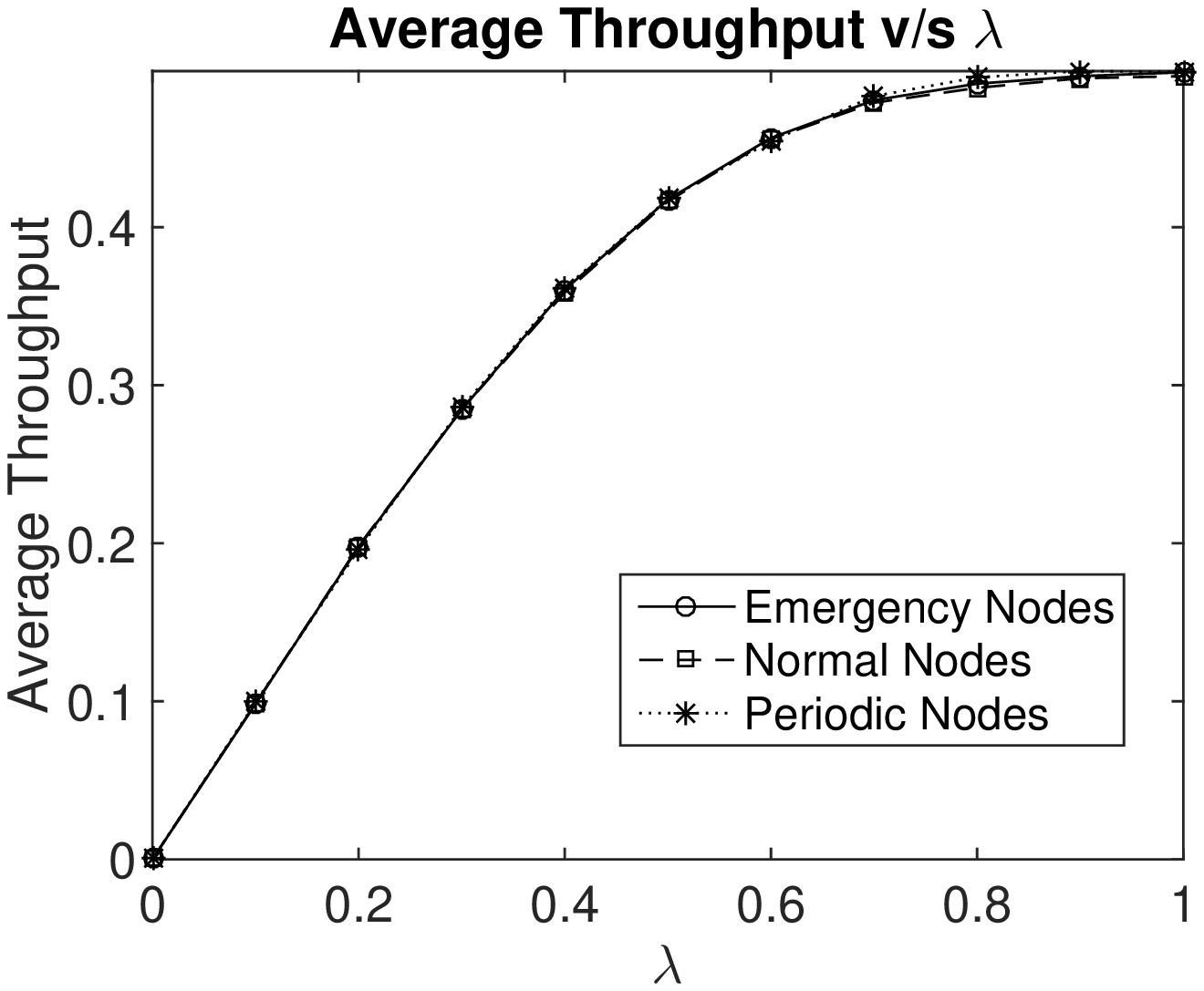}}
    	\label{Avg_Tp_Same_Wt} 
 \end{minipage}%
 ~
\begin{minipage}[b]{0.5\linewidth}
  \centering
  \resizebox{1.0\columnwidth}{!}{\includegraphics{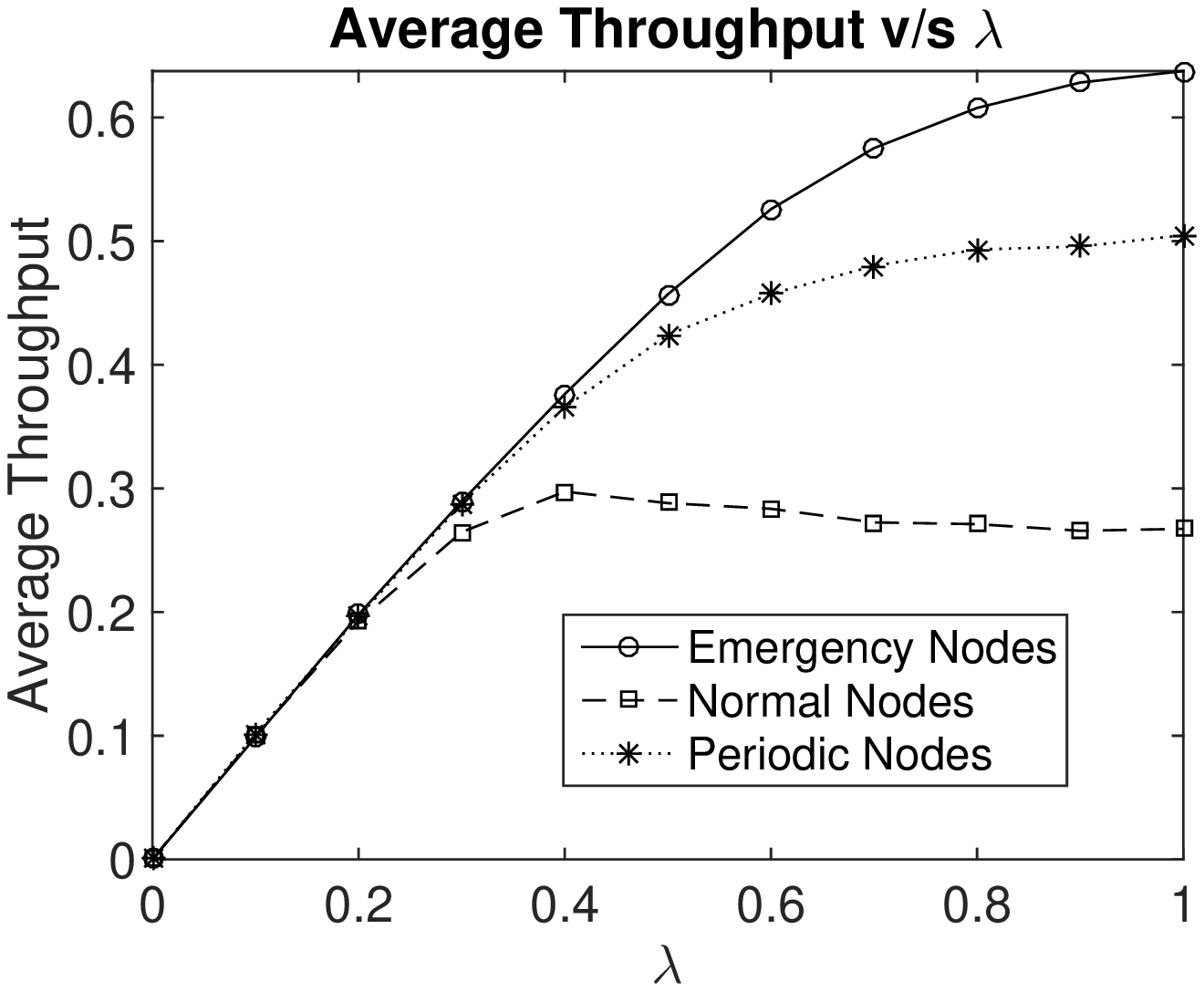}}
    \label{Avg_Tp_Diff_Wt} 
\end{minipage}
    \caption{The following parameters are used in these plots: $M_T = 30$, $N = 50$, $k_{e} = k_{p} = k_{n} = 1$. In the left plot we use $w_{e} = w_{p}= w_{n} = 1$ whereas in the right plot we use $w_{e} = 3$, $w_{p} = 2$, $w_{n} = 1$.}
    \label{AT}
\end{figure}

\begin{figure}[ht]
  \centering  
  \begin{minipage}[b]{0.5\linewidth}
        \centering
        \resizebox{1.0\columnwidth}{!}{\includegraphics{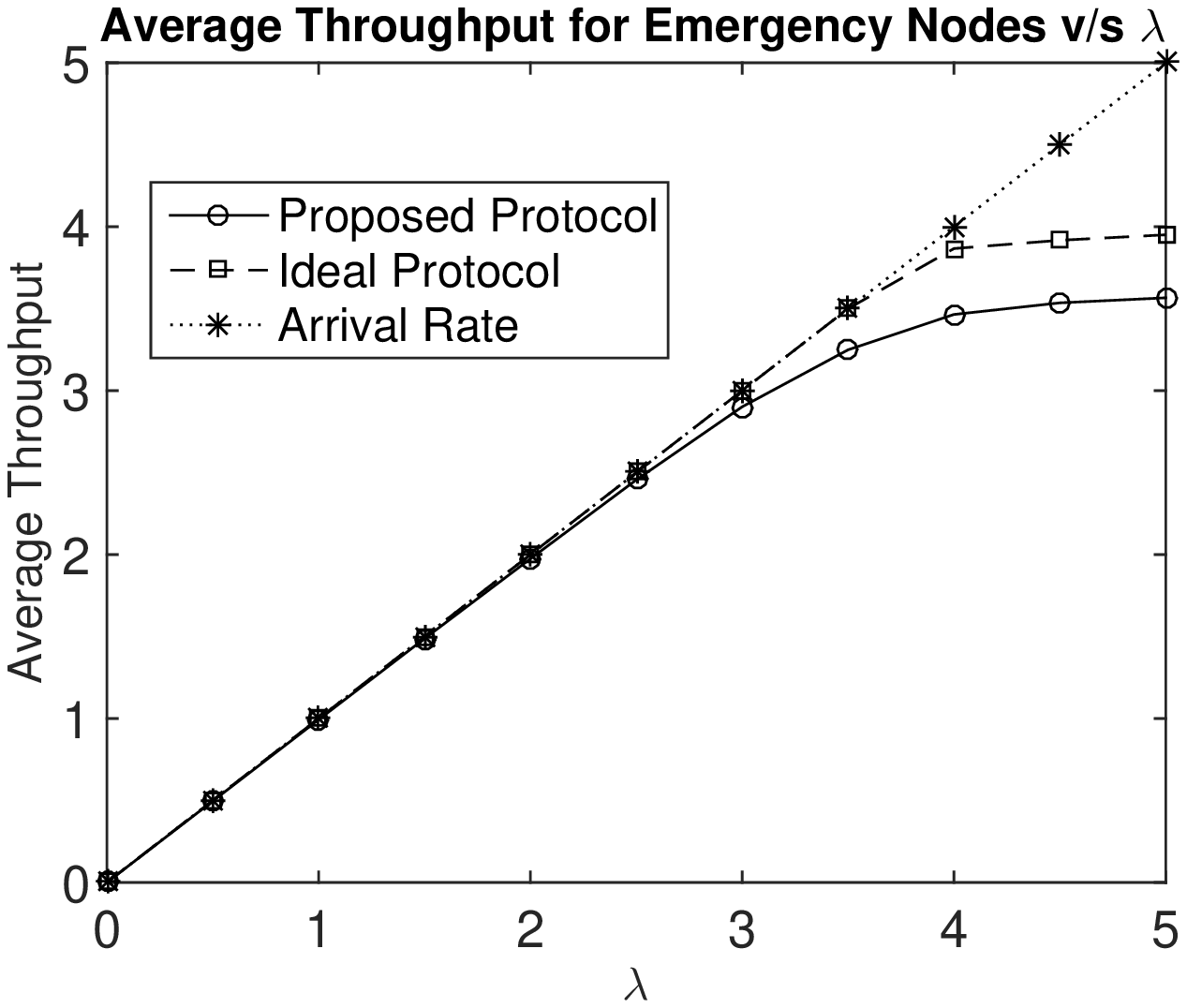}}
    	\label{emergency} 
\end{minipage}%
\begin{minipage}[b]{0.5\linewidth}
  \centering
  \resizebox{1.0\columnwidth}{!}{\includegraphics{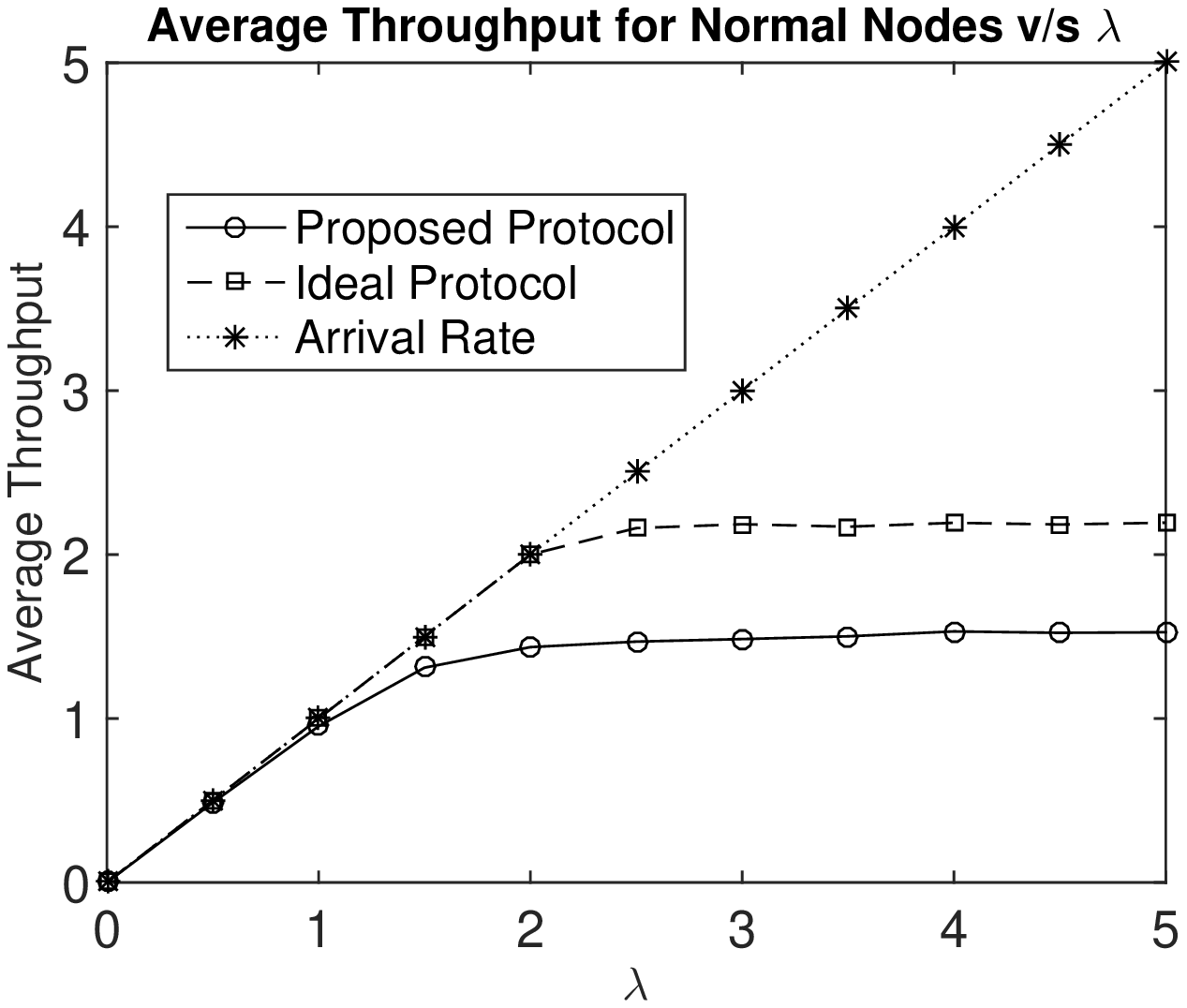}}
     \label{normal} 
\end{minipage}
    \caption{The following parameters are used in these plots: $M_T = 30$, $N = 50$, $w_{e} = 3$, $w_{p} = 2$, $w_{n} = 1$, $k_{e} = k_{p} = k_{n} = 5$.}
    \label{ideal}
\end{figure}

\Figref{AT} shows the average throughput per node versus $\lambda$ for the three classes with different parameter values. It can be seen that initially the average throughput for any given class equals the arrival rate $\lambda$, but after a particular value of $\lambda$, the average throughput saturates, i.e., the system transitions to the unstable region of operation. The left plot in \Figref{AT} shows that when $w_{e}=w_{p}=w_{n}=1$ and $k_{e}=k_{p}=k_{n}=1$, the average throughput curves of all three classes roughly coincide; this is because they are treated alike by the protocol. In contrast, the right plot in \Figref{AT} shows that in the unstable region, emergency (respectively, periodic) nodes achieve a higher average throughput  than  periodic (respectively, normal)  nodes when $w_{e} = 3$, $w_{p} = 2$, $w_{n} = 1$; this is because a higher weight results in more channels being allocated to a class. 
 \Figref{ideal} shows the average throughput  under the proposed protocol and the ideal protocol versus $\lambda$ for the emergency and normal classes. 
These plots show that \emph{the performance of the proposed protocol is close to that of the ideal protocol for both classes}: in particular, in the unstable region of operation, on average, the proposed protocol achieves $87.5\%$ (respectively, $68\%$) of the average throughput under the ideal protocol for the emergency (respectively, normal) class. 

\begin{figure}[ht]
  \centering  
  \begin{minipage}[b]{0.5\linewidth}
        \centering
        \resizebox{1.0\columnwidth}{!}{\includegraphics{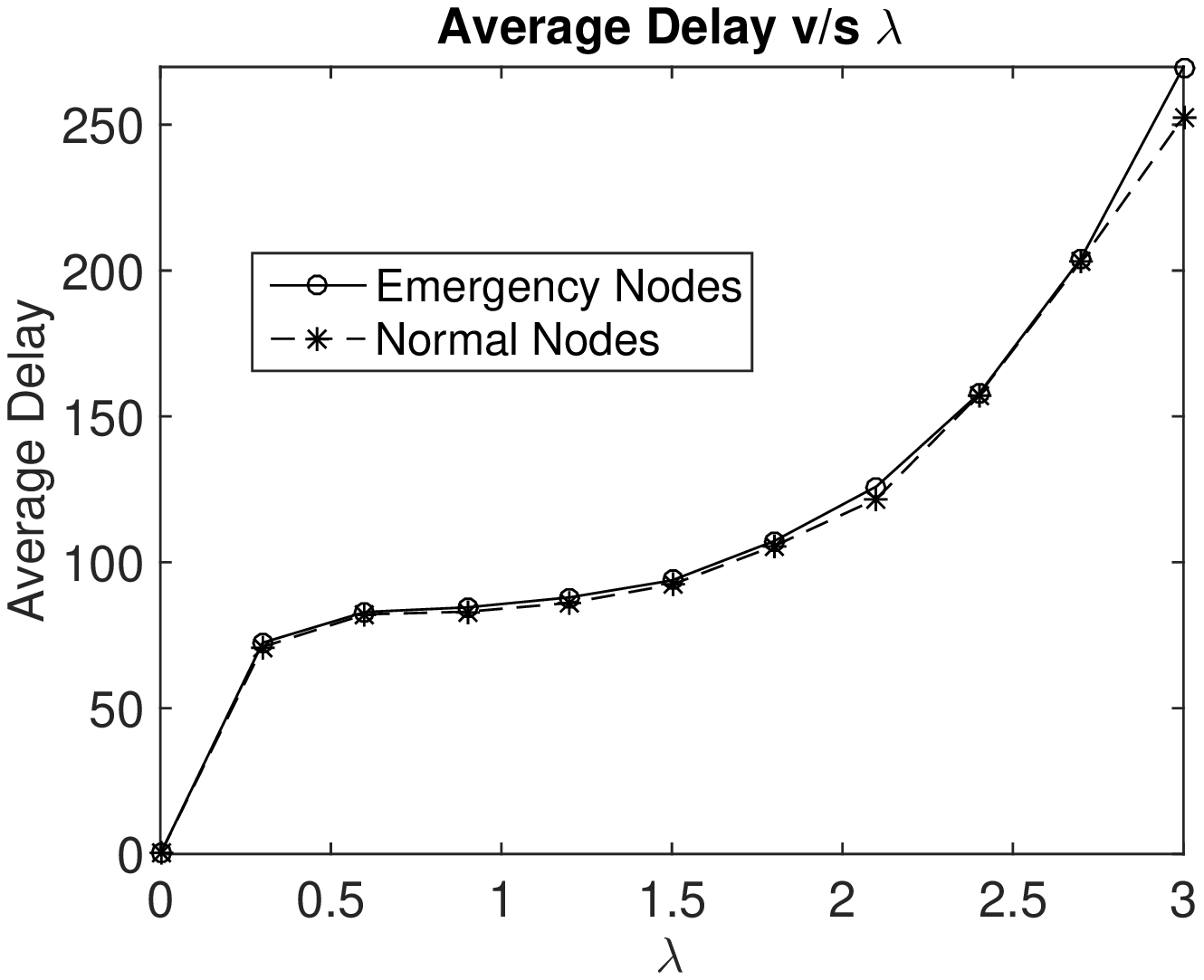}}
    	\label{AD1} 
 \end{minipage}%
\begin{minipage}[b]{0.5\linewidth}
  \centering
  \resizebox{1.0\columnwidth}{!}{\includegraphics{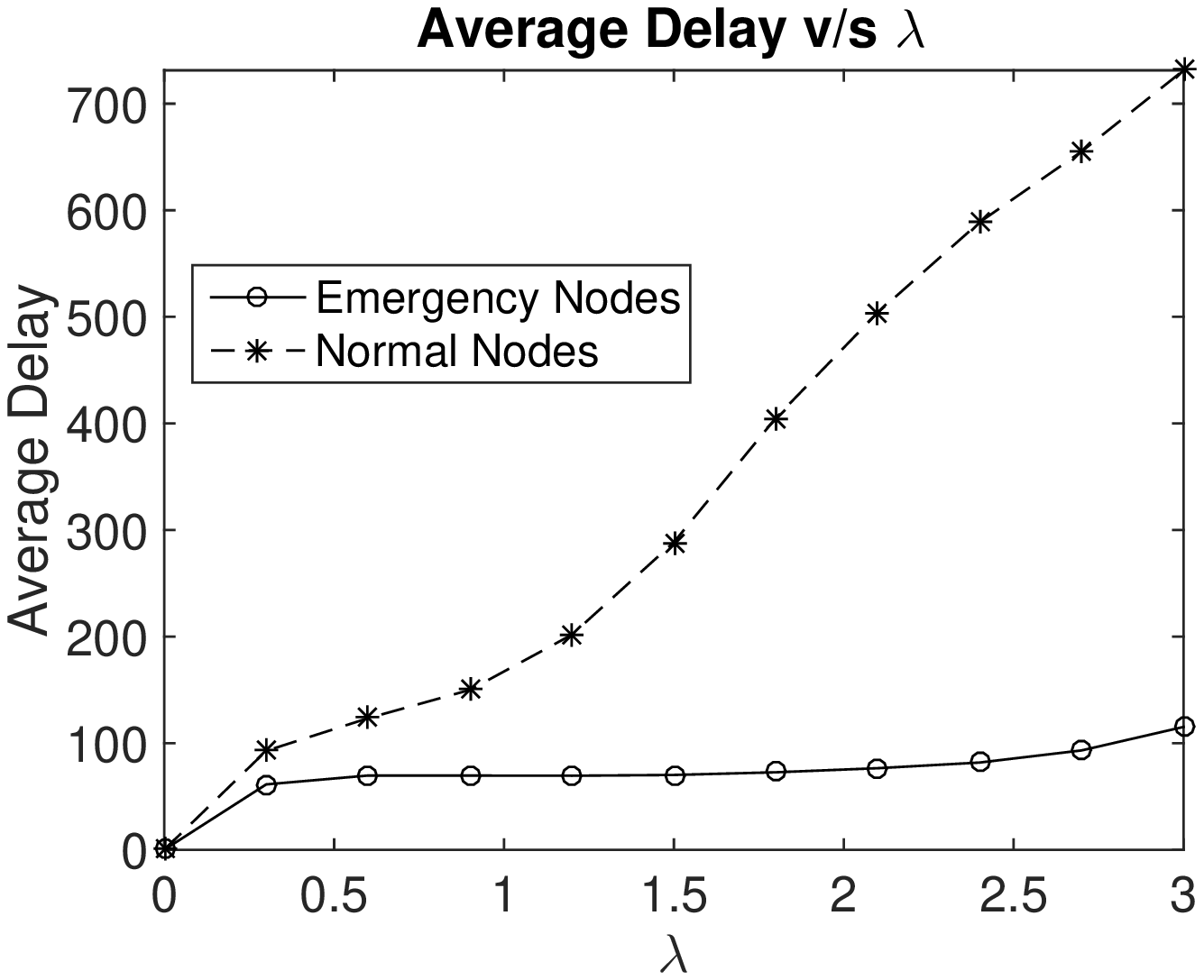}}
    \label{AD3} 
\end{minipage}
    \caption{The following parameters are used in these plots: $M_T = 30$, $N = 50$, $k_{e} = k_{p} = k_{n} =5$. In the left figure we use $w_{e} = w_{p}= w_{n} = 1$ whereas in the right figure we use $w_{e} = 3$, $w_{p} = 2$, $w_{n} = 1$.}
    \label{AD}
\end{figure}

\Figref{AD} shows the average packet delay versus $\lambda$ for the emergency and normal classes with different parameter values. The left plot of \Figref{AD} shows that when $w_{e}=w_{p}=w_{n}=1$ and $k_{e}=k_{p}=k_{n}=5$, the average delay curves of the two classes roughly coincide; on the other hand, when the weights $w_{e} = 3$, $w_{p} = 2$, $w_{n} = 1$ are used (see the right plot of \Figref{AD}), the average delay for emergency nodes is much lower than that of normal nodes, which is because more channels are allocated to emergency nodes.  

\section{Conclusions and Future Work}\label{section:Conclusions}
We  designed two fast node cardinality estimation protocols and a Cognitive MAC protocol for heterogeneous M2M networks. Our MAC protocol uses the proposed fast node cardinality estimation protocols to rapidly estimate the number of active nodes of each type in every time frame; these estimates are used to find the optimal contention probabilities to be used in the MAC protocol. We mathematically analyzed the number of time slots required by one of our proposed estimation protocols (Method I) to execute as well as the performance of the Cognitive MAC protocol. Using simulations, we evaluated the performances of our proposed estimation protocols and Cognitive MAC protocol in terms of the number of time slots required to execute and in terms of the average throughput and average delay achieved respectively.
In this paper, we extended the LoF  based protocol~\cite{qian2011cardinality}, which is an estimation scheme for homogeneous networks, to estimate node cardinalities in heterogeneous networks; a direction for future research is to extend other estimation schemes designed for homogeneous networks in prior work to estimate node cardinalities in heterogeneous networks.   

\bibliography{BibFiles} 
\bibliographystyle{ieeetr}

\appendix

\begin{IEEEproof}[{ Proof of Theorem~\ref{ApproxEKforT}}]
{ Since $K_T = \sum_{i=0}^{t_T-1} I_{\{S_1^{(i)} = C, S_2^{(i)} = C, \ldots , S_{T-1}^{(i)} = C\}}$ (see Section~\ref{secEKERforT}), the expected number of slots required in the second phase is $E(K_T) = \sum_{i=0}^{t_T-1} P(C_i)$, where $C_i$ is the event that collisions occur in all the $T-1$ slots of block $B_i$. Now, $C_i = A_i \cup F_i \cup D_i$, where \\
$A_i$ :  Event that at least two Type 1 nodes transmit in block $B_i$;\\
$F_i$ :  Event that at least two each of Type 2, Type 3, $\ldots$ , Type T nodes transmit in block $B_i$;\\
$D_i$ :  Event that exactly one Type 1 node and one Type 2, one Type 3,  $\ldots$ , one Type T node transmit in block $B_i$.

Using the union bound, upper bounds for $P(A_i)$, $P(F_i)$ and $P(D_i)$ are computed as,
\begin {equation}
\label{PATi}
 P(A_i) \le  {{n_1} \choose {2}} p_i^2 \le \frac{n_1^2}{2}p_i^2; 
\end{equation}

\begin {equation}
P(F_i) \le   {{n_2} \choose {2}} p_i^2 {{n_3} \choose {2}}  p_i^2  \ldots {{n_T} \choose {2}} p_i^2 \le (n_2 n_3 \ldots n_T)^2 \Big(\frac{p_i^2}{2}\Big)^{T-1};
\end{equation}

\begin {equation}
P(D_i) \le (n_1 p_i)(n_2 p_i)\ldots(n_T p_i) = (n_1 n_2 \ldots n_T) p_i^T;
\end{equation}
where $p_i$ is given by \eqref{EQ:pi}. 

Next, recall that $l_{n_r} = \ceil{(\log_2n_r)}$. Using the union bound, $P(C_i) \le P(A_i) + P(F_i) + P(D_i) \le ({n_1^2}/{2})p_i^2 + (n_2 n_3 \ldots n_T)^2 (p_i^2/{2})^{T-1} + n_1 n_2 \ldots n_T p_i^T$. Now, 
\begin{equation}
\label{PCT_Eq}
E(K_T) = \sum_{i=0}^{t_T-1} P(C_i) = \sum_{i=0}^{l_{n_r}  - 2} P(C_i) + \sum_{i= l_{n_r} - 1}^{t_T-1} P(C_i). 
\end{equation}
Let us consider the first summation of \eqref{PCT_Eq}, 
\begin {align}
\label{PCi}
\sum_{i=0}^{l_{n_r}  - 2} P(C_i) \le \sum_{i=0}^{ l_{n_r}  - 2} 1=  l_{n_r}  - 1.
\end{align}
Now consider the second summation of \eqref{PCT_Eq}, in which $t_T$ can be written as $t_T = l_{n_r}  + s$,
\begin{dmath*}
\sum_{i={l_{n_r} - 1}}^{l_{n_r}  + s - 1} P(C_i) \le \sum_{i={l_{n_r} -1}}^{l_{n_r}  + s -1} \left(\frac{n_1^2}{2}p_i^2 + \frac{(n_2 n_3 \ldots n_T)^2}{2^{T-1}} p_i^{2(T-1)} + {n_1 n_2 \ldots n_T} p_i^T\right) \\
\end{dmath*}
\begin{dmath}
= \frac{n_1^2}{2}  \Bigg[\sum_{i={l_{n_r} -1}}^{l_{n_r}  + s -2} \left( \frac{1}{4}\right)^{i+1} + \left( \frac{1}{4}\right)^{l_{n_r}  + s-1} \Bigg]  +  \frac{(n_2 n_3 \ldots n_T)^2}{2^{T-1}}  \Bigg[\sum_{i={l_{n_r} -1}}^{l_{n_r}  + s -2} \left(\frac{1}{4^{T-1}}\right)^{i+1} + \left(\frac{1}{4^{T-1}}\right)^{l_{n_r}  + s-1}\Bigg]  + n_1 n_2 \ldots n_T  \Bigg[\sum_{i={l_{n_r} -1}}^{l_{n_r}  + s - 2} \left(\frac{1}{2^T}\right)^{i+1} + \left(\frac{1}{2^T}\right)^{l_{n_r}  + s-1}\Bigg] \ (\mbox{by }  \eqref{EQ:pi})
\label{PCT_UB}
\end{dmath}

Let us consider the series, $S(r,n) = 1 + r + r^2 + \ldots + r^{n-1} + r^{n-1}$, where $n$ is an integer and $0 < r < 1$. A simplified formula for the sum $S(r,n)$ is as follows.
\begin{equation}
\label{GeoEq}
S(r,n) = \frac{1 - r^n}{1-r} + r^{n-1} = \frac{1-2r^n(1-1/2r)}{1-r}.
\end{equation} 

Let us consider individual upper bounds for each quantity in \eqref{PCT_UB}. We get,
\begin{dmath}
\label{PATi2}
\nonumber \frac{n_1^2}{2}  \Bigg[\sum_{i={l_{n_r} -1}}^{l_{n_r}  + s -2} \left( \frac{1}{4}\right)^{i+1} + \left( \frac{1}{4}\right)^{l_{n_r}  + s-1} \Bigg]   \\   \le  \frac{n_1^2}{2} \left[\frac{1}{n_r^2} + \frac{1}{4n_r^2}  + \frac{1}{16n_r^2} + \ldots \ldots + \frac{1}{4^{s-2}n_r^2} +  \frac{1}{4^{s-2}n_r^2 }\right] \footnotemark     = \frac{n_1^2}{2n_r^2} \Bigg[ \frac{1-2(1/4^s)(1-4/2)}{1-1/4}\Bigg] \text{  \hspace{5mm}    (by~\eqref{GeoEq})}= \frac{2}{3} \frac{n_1^2}{n_r^2}\Big( 1 + 2/4^s\Big).
\end{dmath}
\footnotetext{{ $\left( \frac{1}{4}\right)^{l_{n_r} + k} \le $$ \left( \frac{1}{4}\right)^{\log_2n_r} \left( \frac{1}{4}\right)^k = \left( \frac{1}{4^k n_r^2}\right)$ for $k = 0, 1, 2, \dots, s -2$.}}

\begin{dmath}
\frac{(n_2 n_3 \ldots n_T)^2}{2^{T-1}}  \Bigg[\sum_{i={l_{n_r} -1}}^{l_{n_r}  + s -2} \left(\frac{1}{4^{T-1}}\right)^{i+1} + \left(\frac{1}{4^{T-1}}\right)^{l_{n_r}  + s-1}\Bigg]  \le  \frac{(n_2 n_3 \ldots n_T)^2}{2^{T-1}} \left(\frac{1}{4^{T-1}}\right)^{\log_2n_r} \left[1 + \frac{1}{4^{(T-1)}} + \frac{1}{4^{2(T-1)}}  + \dots +\frac{1}{4^{(s-1)(T-1)}} +\frac{1}{4^{(s-1)(T-1)}}\right] \footnotemark  = \Big(\frac{n_2 n_3 \ldots n_T}{n_r^{T-1}}\Big)^2 \frac{1}{2^{T-1} (1 - 4^{-(T-1)})} \bigg(1 - \frac{2}{4^{(T-1)s}} \Big(1 - \frac{4^{T-1}}{2}\Big)\bigg) \text{ (by~\eqref{GeoEq})}.
\end{dmath}
\footnotetext{{ $\left( \frac{1}{4^{T-1}}\right)^{l_{n_r} + k} \le $$ \left( \frac{1}{4^{T-1}}\right)^{\log_2n_r} \left(  \frac{1}{4^{T-1}}\right)^k $ for $k = 0, 1, 2, \dots, s-2$.}}

\begin{dmath}
\label{UB6}
n_1 n_2 \ldots n_T  \Bigg[\sum_{i={l_{n_r} -1}}^{l_{n_r}  + s - 2} \left(\frac{1}{2^T}\right)^{i+1} + \left(\frac{1}{2^T}\right)^{l_{n_r}  + s-1}\Bigg] \le n_1 n_2 \ldots n_T \left(\frac{1}{2^{T}}\right)^{\log_2n_r} \left[1 + \frac{1}{2^T} + \frac{1}{2^{2T}}  + \dots +  \frac{1}{2^{(s-1)T}} +  \frac{1}{2^{(s-1)T}}\right] \footnotemark = \Big(\frac{n_1 n_2 \ldots n_T}{n_r^T}\Big)  \frac{1}{1 - 2^{-T}} \Big( 1 - \frac{2}{2^{Ts}} (1 - 2^{T-1})\Big) \text{   (by~\eqref{GeoEq})}.
\end{dmath}
\footnotetext{{ $\left( \frac{1}{2^{T}}\right)^{l_{n_r} + k} \le $$ \left( \frac{1}{2^{T}}\right)^{\log_2n_r} \left( \frac{1}{2^T}\right)^k $ for $k = 0, 1, 2, \dots, s-2$.}}
By \eqref{PCT_Eq}--\eqref{PCT_UB} and \eqref{PATi2}--\eqref{UB6}, we get:
\begin{align}
\label{EK_T}
\begin{split}
E(K_T)  \le {}& l_{n_r} - 1+  \frac{2}{3} \frac{n_1^2}{n_r^2}\Big(1+2/4^s\Big)  + \Big(\frac{n_2 n_3 \ldots n_T}{n_r^{T-1}}\Big)^2 \\ & \times  \frac{1}{2^{T-1} (1 - 4^{-(T-1)})} \bigg(1 - \frac{2}{4^{(T-1)s}} \Big(1 - \frac{4^{T-1}}{2}\Big)\bigg) \\ & + \Big(\frac{n_1 n_2 \ldots n_T}{n_r^T}\Big)  \frac{1}{1 - 2^{-T}} \Big( 1 - \frac{2}{2^{Ts}} (1 - 2^{T-1})\Big).
\end{split}
\end{align}}
\end{IEEEproof}

\begin{IEEEproof} [{ Proof of Theorem~\ref{ApproxERforT}}]
{ The expected number of slots required in the third phase is (see Section~\ref{secEKERforT}) $E(R_T) = \sum_{i=0}^{t_T-1} P(A_i)$, where the event $A_i$ is as defined in the proof of Theorem~\ref{ApproxEKforT}. Recall that $l_{n_r} = \ceil{(\log_2n_r)}$. 
From the inequalities  shown in~\eqref{PATi},~\eqref{PCi} and~\eqref{PATi2}, we get: 
\begin{dmath}
\label{ER_T}
E(R_T)  \le l_{n_r}  - 1 +  \frac{2}{3} \frac{n_1^2}{n_r^2}\Big(1+2/4^s\Big).
\end{dmath} }
\end{IEEEproof}

\end{document}